\documentclass[pra,superscriptaddress,11pt]{revtex4}
\usepackage{amssymb}
\usepackage[dvips]{graphicx}
\usepackage[dvips]{color}
\usepackage{epsfig}
\usepackage{dsfont}

\newcommand{\half} {\frac{1}{2}}

\newcommand{\sqnbar} {\sqrt{\bar{n}}}
\newcommand{\nbar} {\bar{n}}
\newcommand{\attractp}{\ket{\psi_{1\,att}^{+}}}
\newcommand{\attractm}{\ket{\psi_{1\,att}^{-}}}
\newcommand{\twoattp}{\ket{\psi_{2\,att}^{+}}}
\newcommand{\twoattm}{\ket{\psi_{2\,att}^{-}}}
\newcommand{\nqattp}{\ket{\psi_{N_q\,att}^{+}}}

\newcommand{\sumn} {\sum_{n=0}^{\infty}}

\newcommand{\ket}[1]{ {\left| #1 \right\rangle} }
\newcommand{\bra}[1]{ {\left\langle #1 \right|} }
\newcommand{\abs}[1]{ {\left| #1 \right|}}
\newcommand{\braket}[2]{ {\left\langle #1  |#2\right\rangle} }
\newcommand{\sfrac}[2]{{\textstyle\frac{#1}{#2}}}

\begin{document}

\title{Dynamics of Entanglement and `Attractor' states in The Tavis-Cummings Model}

\author{C. E. A. Jarvis}\email{catherine.jarvis@bristol.ac.uk}
\affiliation{H H Wills Physics Laboratory, University of Bristol,
Bristol BS8 1TL, United Kingdom}%
\author{D. A. Rodrigues}
\affiliation{School of Physics and Astronomy, University of
Nottingham, Nottingham NG7 2RD, United Kingdom}%
\author {B. L. Gy\"{o}rffy}
\affiliation{H H Wills Physics Laboratory, University of Bristol,
Bristol BS8 1TL, United Kingdom}%
\author {T. P. Spiller}%
\affiliation{Hewlett Packard Laboratories, Filton Road, Bristol,
BS34 8QZ, United Kingdom}
\author{A. J. Short}
\affiliation{DAMPT, Center of Mathematical Sciences, Wilberforce Road, Cambridge, CB3 0WA, United Kingdom}
\author{J. F. Annett}
\affiliation{H H Wills Physics Laboratory, University of Bristol,
Bristol BS8 1TL, United Kingdom}%

\begin{abstract}

We study the time evolution of $N_q$ two-level atoms (or
qubits) interacting with a single mode of the quantised radiation
field. In the case of two qubits, we show that for a set of initial
conditions the reduced density matrix of the atomic system
approaches
 that of a pure state at $\sfrac{t_r}{4}$, halfway between that start of the collapse and the first mini revival peak, where $t_r$
 is the time of the main revival. The pure state approached is the same for a set of initial conditions and is thus termed an `attractor state'.
 The set itself is termed the basin of attraction and the features are at the center of our attention.
 Extending to more qubits, we find that attractors are a generic feature of the multi qubit Jaynes Cummings
 model (JCM) and we therefore generalise the discovery by Gea-Banacloche for the one qubit case. We give the `basin of attraction'
 for $N_q$ qubits and discuss the implications of the `attractor' state in terms of the dynamics of
$N_q$-body entanglement. We observe both collapse
 and revival and sudden birth/death of entanglement depending on
 the initial conditions.
\end{abstract}

\maketitle
\section{Introduction}
The dynamics of two level quantum systems (also known as qubits) such as spins in a magnetic field,
Rydberg atoms or superconducting qubits, coupled to a single mode of an electromagnetic cavity
are of considerable interest in connection with NMR studies of atomic nuclei \cite{Slichter},
cavity quantum electrodynamics \cite{Berman} and the physics of quantum computing \cite{NielsenChng}.
The simplest model that captures the salient features of the physics in these
fields is the Jaynes-Cummings model (JCM) \cite{JaynesCum}, which deals with only one
qubit and its generalisation by Tavis and Cummings to the case of multiple qubits \cite{TavisCum}.

These models have many interesting features \cite{ShoreKnight93}, but perhaps the most
surprising prediction is the phenomena of `collapse and revival' of the Rabi oscillations
in the qubit system when the field is started in a coherent state $\ket{\alpha}$ \cite{GerryKnight},
\begin{equation}\label{eq:coherent}
\ket{\alpha}=e^{-\abs{\alpha}^2/2}\sum_{n=0}^{\infty}
\frac{\alpha^n}{\sqrt{n!}}\ket{n},\ \ \ \ \ \ \alpha=\abs{\alpha}e^{-i\theta}\; .
\end{equation}
Here the state $\ket{n}$ is the eigenstate of the photon number operator $\hat{n}=a^{\dagger}a$
with eigenvalue $n$ and $\theta$ is the initial phase of the radiation field. $\abs{\alpha}$ determines the size of the field, with
$\nbar = \bra{\alpha} \hat{n} \ket{\alpha} = |\alpha|^{2}$.

In short what is found is that the initial Rabi-oscillations of the probability of being in a
given qubit state decay on a time scale called the collapse time, $t_c$, but then revive
after a much longer time, $t_r$, due to the quantum `graininess' of the radiation
field \cite{Cummings,BarnettRadmore,Eberly}. Whilst the nature of this remarkable phenomenon
in the case of one qubit is now well understood \cite{ShoreKnight93}, both experimentally
\cite{Berman} and theoretically \cite{Yoo}, the multi qubit case \cite{TavisCum,ueda} has
not been explored to the same extent. For instance, the occurrence between $t_c$ and $t_r$,
of what we shall call the `attractor' state has only been fully investigated in the one qubit
limit. This intriguing aspect of quantum dynamics was discovered and highlighted by
Gea-Banacloche \cite{Ban90,Ban91} in his study of the one qubit case. In this paper
we report on our investigation of analogous `attractors' in cases of more than one qubit.

The organization of this paper is as follows. In Sec.
\ref{sec:1qubit}, the dynamics of the resonant JCM (single qubit
coupled to a cavity system) is presented. This sets up our notation,
analytic tools and methods of data presentation. In Sec.
\ref{sec:2qubit} we extend the model to two qubits and show that the
idea of an `attractor' state still exists. We investigate the
dynamics in depth and quantify the set of initial states that
approach the `attractor', giving the `basin of attraction'. We show
that the atomic system exhibits interesting entanglement
properties, in particular that the entanglement can vanish and
then return at a later time. Furthermore, the manner in which this
occurs depends on whether we are in or outside this basin of
attraction. In Sec. \ref{sec:nqqubits} we develop these ideas
further from those of the two-qubit case and find a set of initial
states (the basin of attraction) from which the system always goes
to the `attractor' state for an arbitrary number of qubits. In the
limit of large qubit number we rewrite our qubit equations in terms
of spin coherent states and demonstrate some intriguing phenomena
associated with the $N_q$-qubit JCM, especially the collapse
and revival of $N_q$-body entanglement. Whilst our method of
analysis, and some of our results, are similar to those of Chumakov
\emph{et al.} \cite{Chumakov94} and Meunier \emph{et al.}
\cite{Meunier}, our focus is on the quantum information aspects of
the problem and in particular the dynamics of entanglement.

\section{Features of the Jaynes-Cummings Model}\label{sec:1qubit}

We first review the JCM and appearance of `attractor' states in the
one qubit case  \cite{Schrodinger,Buzek,KnightShore}. The JCM model
consists of a single qubit coupled to a single mode of a quantised
electromagnetic field. The two possible states of the qubit are the
ground state $\ket{g}$, with energy $\epsilon_g$, or its excited
state $\ket{e}$, with energy $\epsilon_e$. We consider the following
Hamiltonian:
\begin{eqnarray}\label{eq:1qham}
\hat{H}&=&\hbar\omega\hat{a}^{\dag}\hat{a}+\frac{\hbar\Omega}{2}\hat{\sigma}^{z}+\hbar
\lambda\left(\hat{a}\hat{\sigma}^{+}+\hat{a}^{\dag}\hat{\sigma}^{-}\right)\\
\hat{\sigma}^{z}&=&\ket{e}\bra{e}-\ket{g}\bra{g},\
\hat{\sigma}^{+}=\ket{e}\bra{g},\
\hat{\sigma}^{-}=\ket{g}\bra{e}\label{eq:sigmas}
\end{eqnarray}
where $\hat{a}^{\dag}$ and $\hat{a}$ are the creation and
annihilation operators of photons with frequency $\omega$,
$\hbar\Omega= \epsilon_e-\epsilon_g$ and $\lambda$ the cavity-qubit
coupling constant. The model is only valid close to resonance and we
shall only consider the resonant case of $\omega=\Omega$.

We have an initial state:
\begin{equation}\label{eq:initoneq}
\ket{\Psi_1(t=0)}=\ket{\psi_1}\ket{\alpha}=(C_g\ket{g}+C_e\ket{e})\sumn
C_n \ket{n},
\end{equation}
where the state of the qubit is normalised so
$\abs{C_{e}}^2+\abs{C_{g}}^2=1$ and the field is started in a
coherent state. In the interaction picture (i.e. in a rotating
reference frame) we obtain a solution to the Hamiltonian:
\begin{eqnarray}\label{eq:1qsoln}
\ket{\Psi_1(t)}&=&\sumn \left[
\left(C_eC_n\cos\left(\lambda \sqrt{n+1} t\right)-iC_gC_{n+1}\sin\left(\lambda \sqrt{n+1} t\right)\right)\ket{e,n}\right.\nonumber\\
&&\left.\left(C_gC_{n+1}\cos\left(\lambda \sqrt{n+1}
t\right)-iC_eC_{n}\sin\left(\lambda \sqrt{n+1}
t\right)\right)\ket{g,n+1}\right]+C_gC_0\ket{g,0}.
\end{eqnarray}

In spite of its relative simplicity, Eq. (\ref{eq:1qsoln}) exhibits
a wide range of interesting phenomena. A much studied example is the
collapse and revival of Rabi oscillations. If the qubit is started
in the ground state then initially the probability of the qubit
being in the ground state oscillates from one to zero at the Rabi
frequency. These oscillations then decay on a time scale called the
collapse time, $t_c=\frac{2}{\lambda}$, but then revive after a much
longer time:
\begin{equation}
t_r=\frac{2\pi\sqnbar}{\lambda}.
\end{equation}

When $\nbar\gg \lambda t$ we can approximate $\ket{\Psi_1(t)}$ in Eq.
(\ref{eq:1qsoln}) by $\ket{\widetilde{\Psi}_1(t)}$:
\begin{eqnarray}\label{eq:1qsolnlargen}
\ket{\widetilde{\Psi}_1(t)}&=&\beta_{+\half}(t)\ket{D_{+
\half}(t)}\ket{\Phi_{+\half}(t)}+\beta_{-\half}(t)\ket{D_{-
\half}(t)}\ket{\Phi_{-\half}(t)},
\end{eqnarray}
where $\beta_{\pm\half}(t)$ is a normalisation factor, $\ket{D_{\pm
\half}(t)}$ is a state of the qubits and $\ket{\Phi_{\pm\half}(t)}$
is a state of the field.
\begin{eqnarray}\label{eq:1qsolnlargena}
\beta_{\pm\half}(t)&=&\frac{1}{\sqrt{2}}e^{\pm i\pi\frac {t}{t_r}   \left(\nbar+1\right)}\left(e^{i\theta}C_e\mp C_g \right)\\
\ket{D_{\pm \half}(t)}&=&\frac{1}{\sqrt{2}}\left(e^{-i\theta}\ket{e}\mp e^{\mp i\pi  t/t_r} \ket{g}\right)\\
\ket{\Phi_{\pm\half}(t)}&=&\ket{e^{\pm i \pi t/t_r}\alpha}
\label{eq:1qsolnlargenb}.
\end{eqnarray}

With the solution written in this form it is clear there are two
distinct parts to the wavefunction. In fact the field just consists
of two different coherent states that are out of phase by $\lambda
t/\sqnbar$. In a penetrating study of this model Gea-Banacloche
\cite{Ban91} noted that at a time $\sfrac{t_r}{2}$, in the very
large $\nbar$ approximation, the qubit disentangles itself completely
from the field and the wavefunction factors into a product state of
the radiation field and the qubit.

To make this more apparent we rewrite Eq. (\ref{eq:1qsolnlargena})
-- Eq. (\ref{eq:1qsolnlargenb}) at the time $t=\sfrac{t_r}{2}$.
\begin{eqnarray}\label{eq:1qsolntr2}
\beta_{\pm\half}(\sfrac{t_r}{2})&=&\pm\frac{i}{\sqrt{2}}e^{\pm i \pi \nbar/2}\left(e^{i\theta}C_e\mp C_g \right)\\
\ket{D_{\pm \half}(\sfrac{t_r}{2})}&=&\frac{1}{\sqrt{2}}\left(e^{-i\theta}\ket{e}+i\ket{g}\right)\\
\ket{\Phi_{\pm\half}(\sfrac{t_r}{2})}&=&\ket{\pm i\alpha}.
\end{eqnarray}

With the wavefunction written in this way it can be seen that the
two states $\ket{D_{+\half}(\sfrac{t_r}{2})}$ and
$\ket{D_{-\half}(\sfrac{t_r}{2})}$ are in fact the same at the time
$\sfrac{t_r}{2}$. We label this state $\ket{\psi_{1\,att}^{+}}$:
\begin{equation}\label{eq:oneqattract}
\ket{\psi_{1\,att}^{\pm}}=\frac{1}{\sqrt{2}}\left(e^{-i\theta}\ket{e}\pm
i\ket{g}\right) \; .
\end{equation}

This state is independent of the initial coefficients $C_e$ and
$C_g$ and only depends on the phase $\theta$ of the initial coherent
state, Eq. (\ref{eq:coherent}). Throughout this paper we follow
Phoenix and Knight \cite{PheKnight91} and call this state an
`attractor' state. Note that we use the term `attractor' to
refer to $\ket{\psi^{\pm}_{1\,att}}$ even though in the theory of
non-linear differential equations the term attractor is reserved for
solutions which, when reached, after evolving from an arbitrary
initial condition, stop changing with time. As a consequence the
qubit state can be factorised out of the wavefunction meaning the
qubit and the field are in a product state:
\begin{eqnarray}\label{eq:1qsolntr2a}
\ket{\widetilde{\Psi}_1(\sfrac{t_r}{2})}&=&\frac{\ket{\psi_{1\,att}^{+}}}{\sqrt{2}}\left(ie^{i
\pi \nbar/2}\left( e^{i\theta}C_e-C_g \right)\ket{i\alpha}-ie^{-i\pi
\nbar/2}\left( e^{i\theta}C_e+C_g \right)\ket{-i\alpha}\right).
\end{eqnarray}

It is clear that all the information of the initial state about the
qubit is stored in the radiation field which happens to be in a
``Schr\"{o}dinger cat'' state. At a later time $t=\frac{3t_r}{2}$,
the qubit again disentangles itself completely from the field and
goes to a pure state. The qubit state at this time is orthogonal to
$\ket{\psi_{1\,att}^{+}}$ and we label it $\ket{\psi_{1\,att}^{-}}$.
This remarkable behaviour is most unlike the consequences of a
linear Schr\"{o}dinger equation satisfied by a qubit without the
cavity field or with a classical field and, as was stressed by
Gea-Banacloche, is a natural route to quantum state preparation. All
initial qubit states tend to the same `attractor' state (which is
determined by the initial phase of the field state) and the quantum
information about the initial qubit state has effectively been
swapped out and encoded into the field at the `attractor times'.

Before we move on to describe the results of numerical simulations
we consider the field, Eq. (\ref{eq:1qsolnlargenb}), at two revival
times $t=t_r$ and $t=2t_r$. At both these times the field states
$\ket{\Phi_{\pm\half}(t)}$ become equal, with
$\ket{\Phi_{+\half}(t_r)}=\ket{\Phi_{-\half}(t_r)}=\ket{-\alpha}$
and
$\ket{\Phi_{+\half}(2t_r)}=\ket{\Phi_{-\half}(2t_r)}=\ket{\alpha}$.
As a consequence the system is once again in a product state as the
field can be factorised out of the wavefunction for the whole state.
The information about the initial state of the qubits has been
transferred from the field back to the qubits.

By considering the average number of photons in the field, $\nbar$,
to be large many interesting observations have been made of the JCM.
It was shown that the system repeatedly evolves from a product state
to an entangled state and back again. Although the analytical
equations for large values of $\nbar$ clearly show this behaviour,
we now consider the exact solution $\ket{\Psi_1(t)}$ for modest
$\nbar$ ($\nbar=50$), shown in Eq. (\ref{eq:1qsoln}) and comment on
the results of numerical simulations for the qubit and the photon
sector separately.

\begin{figure}\centering{
\epsfig{file=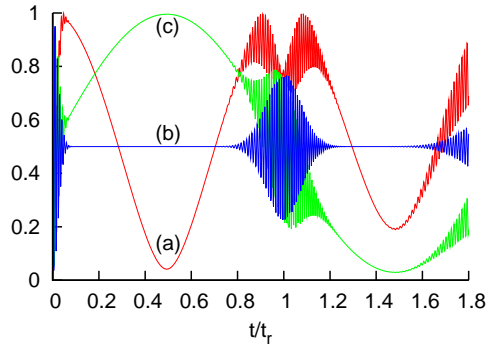, height=5cm}
\caption{(color online) Time evolution for a system with one qubit
when the qubit starts in the initial state $\ket{g}$ and the value
of $\bar{n}=50$. (a) the entropy of the qubit. (b) the probability
of being in the qubit's initial state $\ket{g}$.  (c) the
probability of being in the state $\ket{\psi _{att}^{+}}$. At
$\sfrac{t_r}{2}$ the probability of being in the `attractor' state
goes to one while the entropy goes to zero.\label{fig:oneqexp} }}
\end{figure}

We consider the state of the qubits after tracing out the field
state. The probability of the qubits being in the state
$\ket{g}$ is,
\begin{equation}
P_{1\,g}(t)=\bra{g}\rho_q(t)\ket{g}=\sumn
\abs{\braket{g,n}{\Psi_{1}(t)}}^2
\end{equation}
where
$\rho_q(t)=\mathrm{Tr}_{f}\left(\ket{\Psi_1(t)}\bra{\Psi_1(t)}\right)$
is the reduced density matrix at a certain time for the qubits when
the field has been traced out. Throughout this paper all instances
of the letter $q$ refer to the qubit system and $f$ refers to the
field. The results are depicted in Fig. \ref{fig:oneqexp} for the
initial state $\ket{\Psi_1(t=0)}=\ket{g}\ket{\alpha}$ and the well
known phenomenon of collapse and revival can be seen. For numerical
reasons the sum over $n$ is truncated at $n_{max}=200$ as this is
adequate for the average number of photons in the mode of
$\nbar=50$.

An entropy measurement of the system can be made at any point in
time. Entropy is a well defined quantity that measures the disorder
of a system and the purity of a quantum state. In cases where the
time-dependent Schr\"{o}dinger equation (as opposed to some master
equation for reduced dynamics) determines the time evolution of a
system, the value of the entropy remains constant. The complete
systems (qubits plus field) we study are always started in a pure
state and follow Schr\"{o}dinger dynamics; therefore the entropy of
the complete systems always remains zero as we have a pure state. In
these systems the individual entropies, of the qubits or the field,
are of more interest. We calculate the partial entropy from the
reduced density matrix $\rho_q$ using the rescaled von Neumann
entropy \cite{Neumann,Wehrl},
\begin{equation}\label{eq:vonneu}
S_q(t)=-\mathrm{Tr}(\rho_q(t) \mathrm{log_2} \rho_q(t))/N_q,
\end{equation}
which is rescaled such that the entropy of the qubits ranges from
zero, for a pure state, to one, for a maximally mixed state. For the
bipartite case (the one qubit $N_q=1$ JCM) this reduces to the
standard von Neumann entropy. As was discovered by Araki and Lieb
\cite{Araki70}, if the system is started in a pure state then
$S_q(t)$=$S_f(t)$ for all time, where $S_f(t)$ is the entropy of the
field system calculated from reduced density matrix
$\rho_f(t)=\mathrm{Tr}_{q}\left(\ket{\Psi_1(t)}\bra{\Psi_1(t)}\right)$.
Therefore for the rest of the paper we shall refer to the value of
$S_q(t)$ only when we mention entropy.

In quantum information these partial entropies provide a measure of
the entanglement present between the two systems, known as the
entropy of entanglement \cite{Vedral97,Munro}. If the entropy is
zero then the state being measured is pure and there is no
entanglement present. If the value of the entropy is non zero then
there is some entanglement present between the two parts of the
system and the reduced density matrix is mixed. Returning to the JCM
we note that the `attractor' state, Eq. (\ref{eq:oneqattract}), is a
pure state and hence the value $S_q(\sfrac{t_r}{2})$ will be zero.
So the qubit has disentangled itself from the field and there
is no entanglement present at this time. The reduced entropy is a
powerful tool in understanding entanglement in the JCM and it is
worthwhile recalling an example of its power to elucidate the
quantum dynamics \cite{PheKnight88,PheKnight91}.

We have numerically evaluated $S_q$ for qubits started in the
initial state $\ket{\Psi_1(t=0)}=\ket{g}\ket{\alpha}$ using the
exact solution for $\ket{\Psi_1(t)}$ and the results are shown in
Fig. \ref{fig:oneqexp} by the curve labelled (a). We also plot the
probability
$P_{1\,att}(t)=\bra{\psi^{+}_{1\,att}}\rho_q(t)\ket{\psi^{+}_{1\,att}}$
that the qubit is in the `attractor' state
$\ket{\psi_{1\,att}^{+}}$, as defined in Eq.~(\ref{eq:oneqattract}),
shown by the curve labelled (c).

Evidently at the time $\sfrac{t_r}{2}$ the reduced entropy goes very
close to zero and $P_{1\,att}(\sfrac{t_r}{2})$ goes very close to
one. As mentioned above, this means there is very little
field-qubit entanglement present and the qubit is essentially in the
`attractor' state, $\attractp$, as defined in
Eq.~(\ref{eq:oneqattract}). At $\frac{3t_r}{2}$ the probability of
being in this `attractor' state approaches zero,
 as the qubit is in the state $\ket{\psi_{1\,att}^{-}}$, which is orthogonal to
$\ket{\psi_{1\,att}^{+}}$. These results all agree with the
discussions of Eq. (\ref{eq:1qsolntr2a}). At future `attractor'
times the system alternates between  $\ket{\psi_{1\,att}^{+}}$ and
$\ket{\psi_{1\,att}^{-}}$.

A big difference between numerical results and the large $\nbar$
solution is noticed at the time $t=t_r$. Earlier we showed that the
solution in the large $\nbar$ approximation will return to a pure
state at this time, but in Fig. \ref{fig:oneqexp} the entropy has a
large value of $0.7$, wedged between two maxima (indicating large
entanglement between qubit and field). This is because the width of
this dip is on a time scale similar to the collapse time, much
narrower than the entropy dip at $\sfrac{t_r}{2}$. As the value of
$\nbar$ is increased in numerical simulations, the value of the
entropy at $t=t_r$ gets closer to zero, but we see that even
for fairly large coherent states, the large $\nbar $ approximation can
deviate significantly from the full solution.

So far we have reviewed the time evolution of the reduced qubit
density matrix in the one-qubit and one mode system. Before we move
on to address the two-qubit case we calculate the $Q$ function
\cite{GardinerZoller} to investigate the radiation field,
\begin{equation}\label{eq:qfunc}
Q(\alpha,t)=\bra{\alpha}\rho_f(t)\ket{\alpha}.
\end{equation}

\begin{figure}
\centering{
\epsfig{file=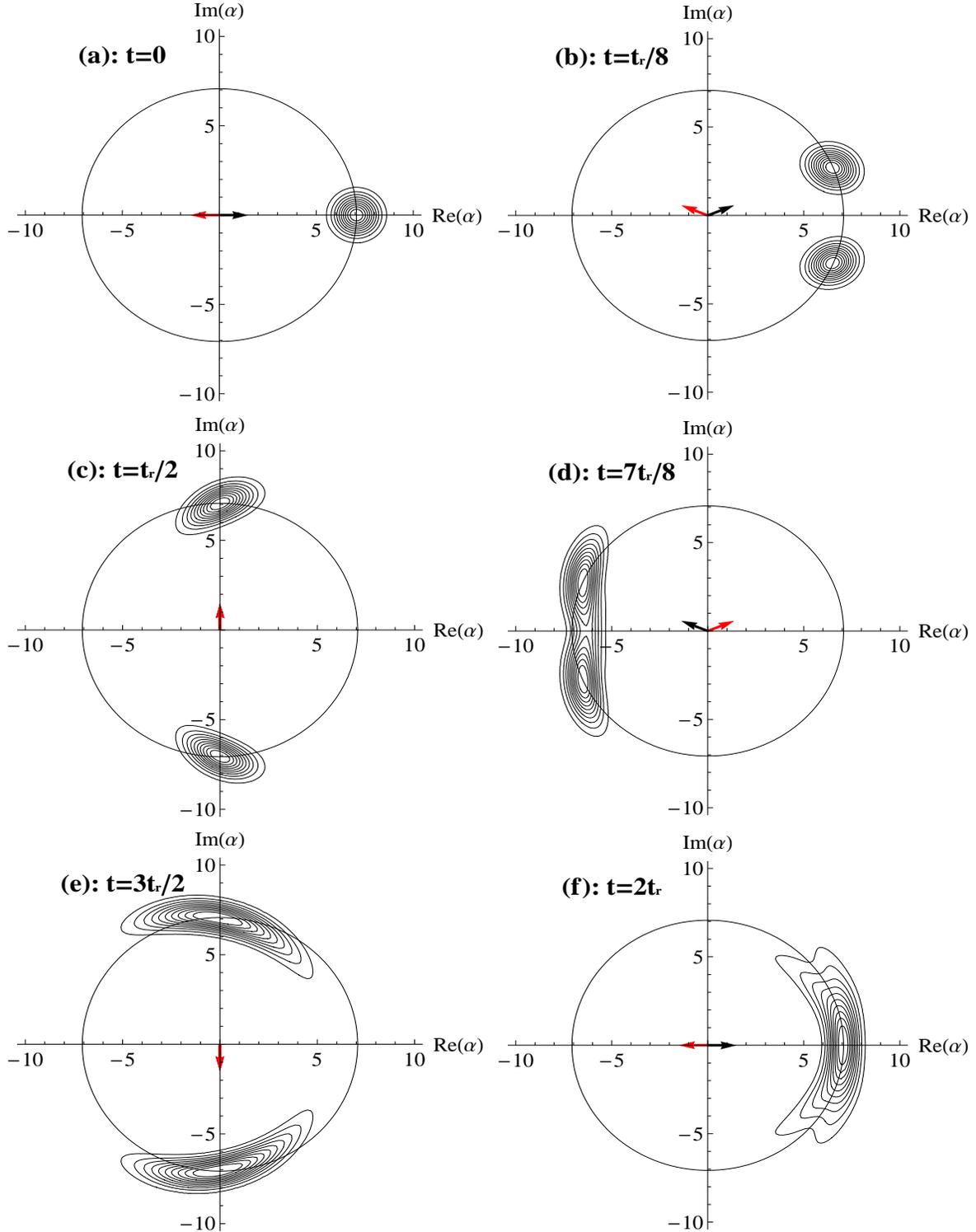, height=20cm}
\caption{(color online) A contour plot of the $Q$ function at six
different times when the qubit is in the initial state $\ket{g}$ and
$\bar{n}=50$. The atomic dipole states are represented by arrows.
(a) the time $t=0$. The system is started in a coherent state which
is shown by a circle of uncertainty in phase space. (b) a time a
little after that of figure (a), $t=\sfrac{t_r}{8}$. The `blobs' are
well separated and are moving around a circle of fixed radius. (c)
the time $t=\sfrac{t_r}{2}$. This corresponds to the first dip in
entropy. (d) just before the time $t=t_r$. This is just before the
revival time and the two `blobs' will overlap. (e) the time
$\sfrac{3t_r}{2}$. The second dip in the entropy. (f) the time
$t=2t_r$. Both the wavepackets have returned to their original
position. \label{fig:1qdipole} }}
\end{figure}

\begin{figure}
\centering{
\epsfig{file=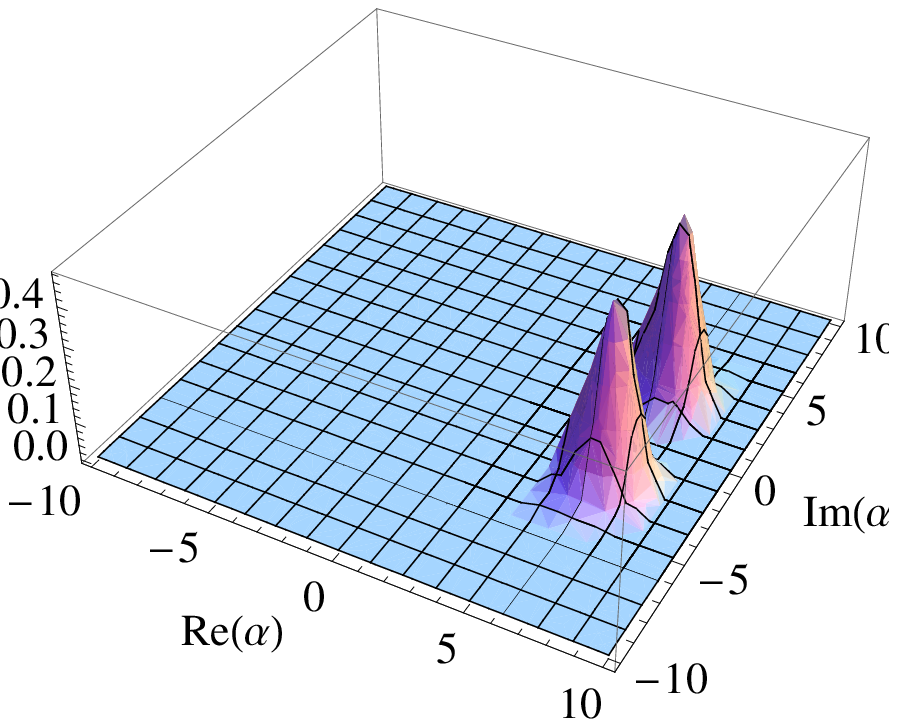, height=7cm}
\caption{(color online) The $Q$ function in 3D for the time
$t=\sfrac{t_r}{8}$ corresponding to diagram (b) in Fig.
(\ref{fig:1qdipole}). We notice two distinct peaks of equal height
that are very similar to coherent states.}\label{fig:1qdipole3d} }
\end{figure}

This is shown in Fig. \ref{fig:1qdipole} in the rotating reference
frame for the exact solution as contour plots. We give an example of
a 3d plot of Fig. \ref{fig:1qdipole}(b) in Fig.
\ref{fig:1qdipole3d}, but for the rest of the paper we will use the
2D contour plots.

The $Q$ function shows a set of two wavepackets of the cavity field,
each of which can be thought of as a Gea-Banacloche state
\cite{Ban91} in the large $\nbar$ solution. As a function of time
the two `blobs' move around the complex plane and follow the
circular path of radius $\sqrt{\bar{n}}$. These `blobs' start to
smear out and become distorted from the original coherent
state. Although the wavepackets all begin in the same place, they
evolve with different frequencies, so the states begin to separate.
After a period of time, depending on the differences between the
frequencies, the different `blobs' are separated by more than their
diameter and can be easily distinguished. Eventually these two
wavepackets will again overlap and this occurs at the revival time,
$t=t_r$. When these 'blobs' have some overlap in phase space there will be Rabi oscillations in the qubit probability, giving the periodic sequence of revivals seen in Fig. \ref{fig:oneqexp}. However, as can be seen in Fig. \ref{fig:1qdipole}(d), at the revival time, $t_r$, the 'blobs' have undergone some distortion and so the wave packets are no longer simple coherent states. This distortion elongates the wave packets, and so they overlap for a longer time than in the ideal coherent state case. We therefore understand the lack of a full amplitude revival, $P_{1g}(t_r)=1$, as due to the spreading, and hence distinguishability, of the field states. When
the value of $\nbar$ gets very large there is less distortion
and the Gea-Banacloche states stay exact coherent states throughout
their evolution.

 Also included in
Fig. \ref{fig:1qdipole} is an arrow for each `blob' which represents
the average atomic polarisation,
$\overrightarrow{d_{\pm\frac{1}{2}}}(t)=\bra{D_{\pm\frac{1}{2}}(t)}
\overrightarrow{\hat{\sigma}} \ket{D_{\pm\frac{1}{2}}(t)}$ in the
equatorial plane of the Bloch sphere. Fig. \ref{fig:1qdipole}(c)
shows the two dipole arrows are in line and pointing in the same
direction at the time $\sfrac{t_r}{2}$ because
$\ket{D_{+\half}(\sfrac{t_r}{2})}=\ket{D_{-\half}(\sfrac{t_r}{2})}=\attractp$.
The arrows in Fig. \ref{fig:1qdipole}(e) are again both the same,
but are pointing in a different direction as
$\ket{D_{+\half}(\frac{3t_r}{2})}=\ket{D_{-\half}(\frac{3t_r}{2})}=\attractm$.
The field states are moving in a different direction in Fig.
\ref{fig:1qdipole}(e) to that of Fig. \ref{fig:1qdipole}(c) and
their identity has swapped. These arrows can be used to see if the
two qubit states are the same and the direction of the dipole arrows
highlights we are in a different `attractor' state.

Inspired by the above behaviour, in this paper we extend the
analysis and investigate new problems. In particular we address the
question: ``Are there similar `attractor' states in multi qubit
systems and what are the implications of their existence?''

\section{Two Qubit Jaynes-Cummings Model}\label{sec:2qubit}

The $N_q$ qubit Hamiltonian, which in this section we treat for
$N_q=2$ so $i=1,2$, is a generalisation of Eq. (\ref{eq:1qham}). It
is called the Tavis-Cummings Hamiltonian \cite{TavisCum}:
\begin{eqnarray}\label{eq:Nqham}
\hat{H}&=&\hbar\omega\hat{a}^{\dag}\hat{a}+\frac{\hbar}{2}\sum_{i=1}^{N_q}\Omega_i\hat{\sigma}_{i}^{z}+\hbar
\sum_{i=1}^{N_q}
\lambda_{i}\left(\hat{a}\hat{\sigma}_{i}^{+}+\hat{a}^{\dag}\hat{\sigma}_{i}^{-}\right).
\end{eqnarray}
For brevity we only consider the solution for the case
$\lambda_1=\lambda_2=\lambda$ and $\omega=\Omega_1=\Omega_2$. This
model has been studied in detail by Chumakov \emph{et al.}
\cite{Chumakov95} and the large $\nbar$ solution has been found
\cite{Chumakov94,Meunier} with similar results to our analysis. Adding another qubit
to the system yields a more complex system which can exhibit
interesting entanglement properties as not only can the two qubits
be entangled with the field, but they can also be entangled between
themselves.

We have investigated the evolution of the system over time, with
analytical solutions in the large $\nbar$ approximation
\cite{RodJar} and the numerically evaluated exact solution. To summarise these we start with the initial state:
\begin{equation}\label{eq:init}
\ket{\Psi_2(t=0)}=\ket{\psi_2}\ket{\alpha}=\left(C_{ee}\ket{ee}+C_{eg}\ket{eg}+C_{ge}\ket{ge}+C_{gg}\ket{gg}\right)\ket{\alpha},
\end{equation}
where the state is normalised so
$\abs{C_{ee}}^2+\abs{C_{eg}}^2+\abs{C_{ge}}^2+\abs{C_{gg}}^2 =1$.
Following the same procedure as in the one qubit case we solve the
Hamiltonian and approximate for large $\nbar$. The time evolution of
this state takes the following form:
\begin{equation}
\ket{\widetilde{\Psi}_2(t)}=\sum^{1}_{k=-1}\beta_{k}(t)\ket{D_{k}(t)}\otimes\ket{\Phi_k(t)},
\end{equation}
where $k= -1, 0$ or $1$ and:
\begin{eqnarray}\label{eq:stateafinal}
\beta_{\pm1}(t)&=&\frac{e^{\pm 2i\pi \frac{t}{t_r}\left(\nbar+2\right)}}{2}(e^{2i\theta}C_{ee}\mp e^{i\theta}(C_{eg}+C_{ge})+C_{gg}),\\
\beta_{0}&=&\frac{1}{\sqrt{2}}\sqrt{\abs{e^{2 i \theta} C_{ee}-C_{gg}}^2+\abs{C_{eg}-C_{ge}}^2},\label{eq:statecfinal}\\
\ket{D_{0}(t)}&=&\ket{D_0}=\frac{(e^{2i\theta}C_{ee}-C_{gg})
(e^{-2i\theta }\ket{ee}-\ket{gg})+(C_{eg}-C_{ge})(\ket{eg}-\ket{ge})}{\sqrt{2\left(\abs{e^{2 i\theta}C_{ee}-C_{gg}}^2+\abs{C_{eg}-C_{ge}}^2\right)}},\\
\ket{D_{\pm 1}(t)} &=&\half[e^{-2i\theta}\ket{ee}
\mp e^{\mp 2i\pi t/t_r}e^{-i\theta}(\ket{eg}+\ket{ge})+e^{\mp 4i\pi t/t_r}\ket{gg}],\label{eq:statebfinal}\\
&=&\ket{D_{\pm \half}\left(2t\right)} \otimes \ket{D_{\pm \half}\left(2t\right)}, \label{eq:2qlargenin1q}\\
\ket{\Phi_k(t)}&=&\ket{e^{2i k\pi
t/t_r}\alpha}.\label{eq:stateradfinal}
\end{eqnarray}
Note that the state $\ket{D_{0}(t)}=\ket{D_{0}}$ is time-independent
and is thus a maximally entangled qubit state for all times if
$C_{eg}=C_{ge}$ \cite{RodJar}. The field states are called
Gea-Banacloche states and once again have a specific qubit state
associated to each of them.

By making the observation shown in Eq. (\ref{eq:2qlargenin1q}) it is
easy to see that the two qubit states $\ket{D_{\pm 1}(t)}$  are
direct products of the one qubit states $\ket{D_{\pm \half}(t)}$ up
to a mapping $t\rightarrow 2t$. This can be used to predict the
behaviour of $\ket{D_{\pm 1}(t)}$. As stated in Sec.
\ref{sec:1qubit} $\ket{D_{\pm \half}(\sfrac{t_r}{2})}=\attractp$
therefore we can conclude that $\ket{D_{\pm 1}(t)}$ will go to a
direct product of $\attractp$ at $t=\sfrac{t_r}{4}$.
\begin{eqnarray}
\ket{D_{\pm1}(\sfrac{t_r}{4})} &=&\ket{D_{\pm \half}\left(\sfrac{t_r}{2}\right)} \otimes \ket{D_{\pm \half}\left(\sfrac{t_r}{2}\right)}\\
&=&\attractp\otimes\attractp\\
&=&\ket{\psi^{+}_{2\,att}}
\end{eqnarray}

So we have found a term that is similar to the one qubit `attractor'
state, $\attractp$, which we shall label the two qubit `attractor'
state, $\twoattp$:
\begin{equation}
\ket{\psi^{\pm}_{2\,att}}=\half\left(e^{2i\theta}\ket{ee}\pm ie^{i\theta}\left(\ket{eg}+\ket{ge}\right)-\ket{gg}\right)
\end{equation}
In the one qubit case the qubit system went to a second `attractor'
state $\attractm$ at the time $\frac{3t_r}{2}$ and means
$\ket{D_{\pm
1}(\frac{3t_r}{4})}=\attractm\otimes\attractm=\ket{\psi^{-}_{2\,att}}$.

\subsection{Basin of Attraction}\label{sec:2qbasin}

The two qubit JCM is more complicated than the original one qubit
case and shows some very interesting properties. Firstly we
rewrite the wavefunction for the whole system as the time
$t=\sfrac{t_r}{4}$ to understand the two qubit case in more detail.
\begin{eqnarray}
\ket{\widetilde{\Psi}_2(\sfrac{t_r}{4})}&=&-\half\ket{\psi_{2\,att}^{+}}\left[\left(e^{2i\theta}C_{ee}+C_{gg}\right)\left(e^{i\pi\nbar/2
}\ket{i\alpha}+e^{-i\pi\nbar/2 }\ket{-i\alpha}\right)\right.
\nonumber\\
&-&\left.e^{i\theta}\left(C_{eg}+C_{ge}\right)\left(e^{i\pi\nbar/2
}\ket{i\alpha}-e^{-i\pi\nbar/2 }\ket{-i\alpha}\right)\right]
+\beta_0\ket{D_{0}}\ket{\alpha} \label{eq:2qwavefunction}
\end{eqnarray}

With the wavefunction written in this form we can see which initial
conditions result in the entropy going to zero. Indeed when
$\beta_0=0$ the qubit and field parts are each in a product state,
so there is no qubit-resonator entanglement present. There is also
no qubit-qubit entanglement as these are in the unentangled
`attractor' state and this means there is absolutely no entanglement
in the system at $t=\sfrac{t_r}{4}$. Interestingly, the information
on the initial state of the qubits is again stored in the radiation
field and the qubits are in a spin coherent `attractor' state (see
appendix \ref{ap:spin}).

The initial conditions of the qubit that results in $\beta_0=0$ are:
$e^{i\theta}C_{ee}=e^{-i\theta}C_{gg}=a$ and
$C_{eg}=C_{ge}=\sqrt{\half -\abs{a}^2}$. So the initial state of
the qubits will be of the form:
\begin{equation}\label{eq:initattract}
\ket{\psi_2}=a(e^{-i\theta}\ket{ee}+e^{i\theta}\ket{gg})+\sqrt{\half-\abs{a}^2}(\ket{eg}+\ket{ge})
\end{equation}
where $\theta$ is the initial phase of the radiation field and
$0\leq\abs{a}\leq\frac{1}{\sqrt{2}}$. These states lie in the
symmetric subspace and have some interesting properties that will be
discussed in section \ref{sec:2qcollapseent}. While the qubit state
is within this basin of attraction its state is parameterized
by the single complex variable $a$ and the wavefunction only
contains two field states, $\ket{\Phi_{\pm 1}(t)}$. At the time
$\sfrac{t_r}{4}$ these two field states are macroscopically distinct
coherent states:
\begin{eqnarray}
\ket{\widetilde{\Psi}_2(\sfrac{t_r}{4})}&=&-e^{i\theta}\ket{\psi_{2\,att}^{+}}\left[e^{i\pi \nbar/2}\left(a-\sqrt{\half-\abs{a}^2}\right)\ket{i\alpha}+ e^{-i\pi \nbar/2}\left(a+\sqrt{\half-\abs{a}^2}\right)\ket{-i\alpha}\right].\nonumber\\
\end{eqnarray}

We now see that the initial state of the qubits alters the relative
proportions of $\ket{\alpha}$ with respect to $\ket{-\alpha}$. This
is also seen in the one qubit case and is the mechanism whereby the
quantum information of the initial qubit state is encoded in the
field. State preparation of a Schr\"{o}dinger cat state utilizing
this phenomenon has been discussed \cite{Ban91,Yurke}. With the two
qubit case the same protocol can be used, but is not so useful as
the initial qubit state has to be within the basin of attraction in
contrast to the one qubit case where all initial qubit states lead
to a Schr\"{o}dinger cat state at the `attractor' times.

There is also a connection between the field states $\ket{\Phi_{\pm
1}(t)}$ in the two qubit case and the field states $\ket{\Phi_{\pm
\half}(t)}$ from the one qubit case, $\ket{\Phi_{\pm
1}(t)}=\ket{\Phi_{\pm \half}(2t)}$. Again there is a mapping
$t\rightarrow 2t$. In Sec. \ref{sec:1qubit} it was shown that the
field and qubit become unentangled at $t_r$ and $2t_r$ as
$\ket{\Phi_{+\half}(t_r)}=\ket{\Phi_{-\half}(t_r)}$ and
$\ket{\Phi_{+\half}(2t_r)}=\ket{\Phi_{-\half}(2t_r)}$. Using this
information from the one qubit case we can predict that
$\ket{\Phi_{+ 1}(\sfrac{t_r}{2})}=\ket{\Phi_{-1}(\sfrac{t_r}{2})}$
and $\ket{\Phi_{+ 1}(t_r)}=\ket{\Phi_{-1}(t_r)}$ which is indeed the
case. As a consequence the field part of the wavefunction can be
factorised out and there is no entanglement present between the
qubits and the field at the times $\sfrac{t_r}{2}$ and $t_r$.

\begin{figure}[tp!]
\centering{
\epsfig{file=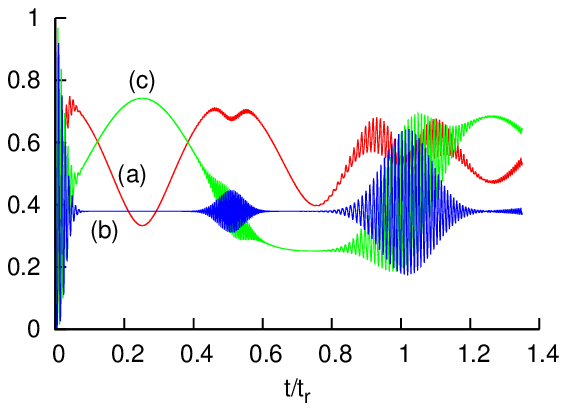,height=8cm}\\
\epsfig{file=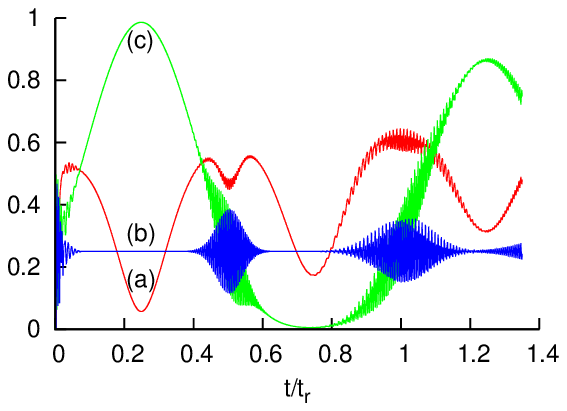,height=8cm}
} \caption{(color online) Time evolution for a system with two
qubits when $\nbar=50$ for two different initial states of the
qubits. (Top)The initial state of the qubits is $\ket{gg}$ and it
can be seen that the two qubit `attractor' state is not reached at
$\sfrac{t_r}{4}$. (Bottom) The initial state of the qubits is
$\frac{1}{\sqrt{2}}(\ket{ee}+\ket{gg})$ (i.e. within the basin
of attraction) and it shows that the two qubit `attractor'
state is reached at $\sfrac{t_{r}}{4}$.  \\
In both diagrams (a) the entropy of the qubits. (b) the probability
of the two qubit state $\ket{gg}$. (c) the probability of being in
the two qubit `attractor' state $\ket{\psi_{2\,att}^{+}}$ when the
initial phase of the radiation field is $\theta
=0$.\label{fig:twoqubitatt}}
\end{figure}

We now consider the numerical results to test the analytical
predictions and see how the system behaves when $\nbar=50$, shown in Fig~\ref{fig:twoqubitatt}. We consider
the probability of the qubits being in the state $\ket{gg}$ and the
probability $P_{2\,att}(t)=\bra{\psi_{2\,att} ^{+}}
\rho_{q}(t)\ket{\psi_{2\,att}^{+}}$, that $\twoattp$ occurs at a
particular time for two sets of initial conditions, one
satisfying $\beta_0=0$ (Fig~\ref{fig:twoqubitatt}(b)) and one not
(Fig~\ref{fig:twoqubitatt}(a)). Evidently the pattern of collapse
and revival is more complicated than in the one qubit case
\cite{Iqbal}, but nevertheless we still see characteristic dips in
the entropy. The value of the entropy at $\sfrac{t_r}{4}$ is not the
same in the two cases shown in Fig. \ref{fig:twoqubitatt} and in
fact they show two completely different behaviours. In Fig.
\ref{fig:twoqubitatt}(a) we notice that the dip in the entropy goes
to a value of 0.35, very different to the value of the entropy in
Fig.~\ref{fig:oneqexp}, but in Fig~\ref{fig:twoqubitatt}(b) we do
indeed find that the entropy is very small at $\sfrac{t_r}{4}$. The
initial state used for Fig. \ref{fig:twoqubitatt}(a) is not in the
basin of attraction, whereas the initial state used for
Fig~\ref{fig:twoqubitatt}(b) is.

A second dip in the entropy can be seen in
Fig.~\ref{fig:twoqubitatt} at the time $t=\frac{3t_r}{4}$ which, in
the large $\nbar$ approximation, goes to zero. The probability of
being in the `attractor' state $\ket{\psi^{+}_{2\,att}}$ goes to
zero at this time. The field has disentangled itself from the field
again, so at this point the qubits are again in a pure state, the
orthogonal `attractor' state $\attractm$.

It can be seen from the analytic solution that, in the large $\nbar$
approximation, the field and qubits also disentangle themselves at the
time $t_r$, for all initial conditions of the qubits, and also at
$\sfrac{t_r}{2}$ if the initial qubit state is in the basin of
attraction. This is not seen in the diagrams in
Fig.~\ref{fig:twoqubitatt} and instead the value to the entropies at
these times are fairly high which means the numerically evaluated exact
system is far from the analytic
results in large $\nbar$ approximation. This is similar
to the one qubit case and can be explained by considering the
dynamics of the field.

\begin{figure}[tp]
\centering{
\epsfig{file=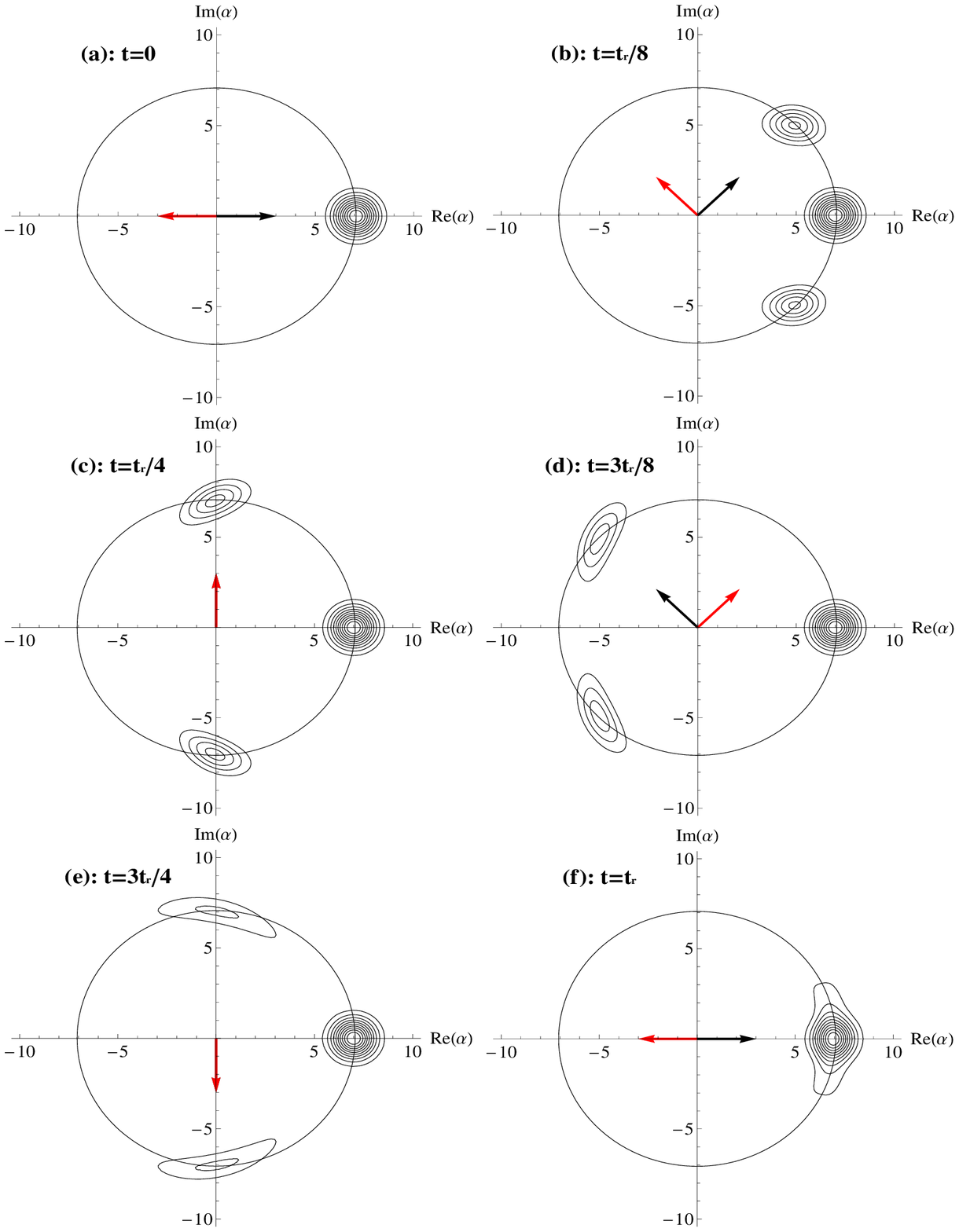,height=20cm}
\caption{(color online) The $Q$ function at six different times when
the qubits are initially in the state $\ket{gg}$. Once again the
atomic dipole states are represented as arrows and $\nbar=50$. (a)
the time $t=0$. The system is started in a coherent state which is
shown by a circle of uncertainty in phase space. (b) a time a little
after that of figure (a), $t=\sfrac{t_r}{8}$. (c) the time
$t=\sfrac{t_r}{4}$. (d) the time $t=\frac{3t_r}{8}$. (e) the time
$\frac{3t_r}{4}$. (f) the time $t=t_r$. Both the wavepackets have
returned to their original position. \label{fig:2qdipole} }  }
\end{figure}

We can once again plot the $Q$ function for the numerical
solution to pictorially display what is occurring in the cavity
field. The values of $Q(\alpha,t)$ in the complex $\alpha$
plane at fixed times are shown in Fig. \ref{fig:2qdipole} when the
qubits are initially in the state $\ket{gg}$, which is outside the
basin of attraction. This time the $Q$ function shows a set of three
wavepackets of the cavity field (compared to the two for the
one-qubit case, Fig. \ref{fig:1qdipole}). As a function of time the
three `blobs' separate, but only two move around the complex plane
and follow the circular path of radius $\sqnbar$. Again the arrows
show the dipole moment of the qubit state associated with each
wavepacket, $\overrightarrow{d_{k}}(t)=\bra{D_{k}(t)}
\overrightarrow{\hat{J_2}} \ket{D_{k}(t)}$. The `blob' that does not
move over time has a dipole moment of zero, so only the two moving
Gea-Banacloche states have an arrow.

For the one-qubit case the revival time occurred when the two
wavepackets overlapped. In this case we have two `blobs' that move
around phase space and overlap at the time $t=\sfrac{t_r}{2}$,
leaving the stationary wavepacket at the initial position. This
causes a rather weak revival peak as only two of the three
wavepackets are in phase. At a later time, Fig.
\ref{fig:2qdipole}(f), all three wavepackets converge at the
starting position and there is a main revival peak, $t=t_r$. If the
initial state of the qubits was in the basin of attraction then
$\beta_0\approx0$ and there would only be two wavepackets. The $Q$
function pictures would then look similar to those in Fig.
\ref{fig:1qdipole} and the system would behave like the
one-qubit case. Nevertheless we note that the radiation field
returns to the start position after one revival time, $t_r$ compared
to the one-qubit case where it returns to the same place after
$2t_r$.

In summary when there are two qubits interacting with a single
cavity mode the wavefunction consists of a part that behaves like
the one qubit case, $\ket{D_{\pm1}(t)}$ and an extra part
$\ket{D_0}$. The `attractor' state, $\ket{\psi_{2\,att}^{\pm}}$,
occurs at the times when the two `blobs' in phase space are
macroscopically different and when $\beta_0=0$. Moving from the one
qubit case to the two qubit case involves a mapping of $t\rightarrow
2t$ and a similar mapping can be used to understand the
equations for more qubits.

\subsection{Collapse and Revival of Entanglement}\label{sec:2qcollapseent}

An interesting new feature of the two-qubit case, as opposed to the
one-qubit case, is that there can be entanglement present between
the two qubits themselves, not just between the qubits and the
field. So far we have used the entropy to measure the amount of
entanglement between the field and the qubit system as the whole
system is in a pure state. Unfortunately this measure can not be
used to calculate the entanglement between the two qubits as they
are sometimes in a mixed state themselves. To find
the amount of entanglement between the two qubits we use the mixed
state tangle $\tau$, defined as \cite{Wooters98,Wooters01}:
\begin{equation}
\tau=[\mathrm{max}\{\lambda_1-\lambda_2-\lambda_3-\lambda_4,0\}]^2.
\label{eq:mixedtangle}
\end{equation}
The $\lambda's$ are the square root of the eigenvalues, in
decreasing order, of the matrix
$\rho_{q}\left(\sigma^y\otimes\sigma^y\right)\rho_{q}^{\ast}\left(\sigma^y\otimes\sigma^y\right)$,
where $\sigma^y=i\left(\sigma^{-}-\sigma^{+}\right)$ (see equation
Eq. \ref{eq:sigmas}) and $\rho_{q}^{\ast}$ represents the complex
conjugate of $\rho_{q}$ in the $\ket{e}$, $\ket{g}$ basis. A tangle
of unity indicates that there is maximum entanglement between the
two qubits. If the tangle takes the value of zero, then there is no
entanglement present between the two qubits. To clarify, the entropy
will be used to measure the amount of entanglement between the field
and the qubit system whereas the tangle will be used to measure the
entanglement between the qubits.

\begin{figure}[h!]
\centering{
\epsfig{file=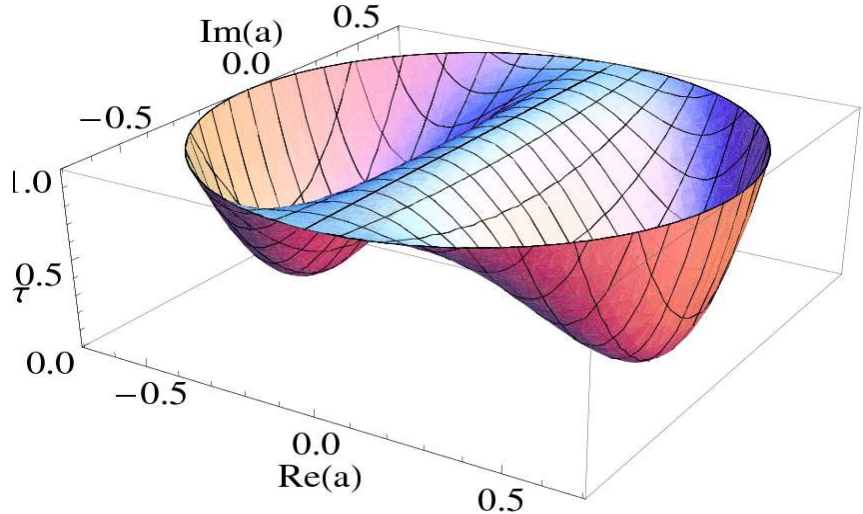,height=7cm}
\caption{(color online) The value of the tangle for the states in
the basin of attraction, for different values of $a$. We notice that
the plot shows only two points where the tangle is zero.}
\label{fig:concur}}
\end{figure}

In this subsection we focus on the initial states inside the basin
of attraction. The value of the tangle for the states given by
Eq.~(\ref{eq:initattract}) can be seen in Fig.~\ref{fig:concur}.
This diagram shows only two points where $\tau=0$, for $a=\pm
\half$, which demonstrates that there are only two product states in
the `basin of attraction'. In contrast, there are many points in
which $\tau=1$, for $a=\frac{e^{i\phi}}{\sqrt{2}}$ where $\phi$ is
an arbitrary phase, or $a=ir$ where $r$ is a real number. This means
that all possible levels of entanglement are represented in the
basin of attraction, Eq.~(\ref{eq:initattract}).

Whilst almost all states in the basin of attraction given in
Eq.~(\ref{eq:initattract}) describe entangled states, these all
evolve into $\twoattp$ at $t=\sfrac{t_r}{4}$ and $\twoattm$ at
$t=\frac{3t_r}{4}$, at which times they are pure and unentangled. So
if the system is started off in one of these entangled states
defined by Eq. (\ref{eq:initattract}), at the time $\sfrac{t_r}{4}$
the system goes to a state which has zero entanglement. Therefore we
have gone from having entanglement in the system to having
absolutely no entanglement, either between the qubits and the
resonator or between the qubits themselves. At a later time this
entanglement returns to the system and can be said to `revive'. The
tangle returns to its initial value at $\sfrac{t_r}{2}$ and
$t_r$, regardless of the phase of the coherent state. This can
be seen when we consider the wavefunction for the system at this
time:
\begin{eqnarray}
\ket{\widetilde{\Psi}_2(\sfrac{t_r}{2})}&=&-\frac{1}{2}\ket{-\alpha}\left[e^{i\pi\nbar}\left(a-\sqrt{\half-\abs{a}^2}\right)\left(e^{-i\theta}\ket{ee}+\ket{eg}+\ket{ge}+e^{i\theta}\ket{gg}\right)\right.\nonumber\\
&+&\left.e^{-i\pi\nbar}\left(a+\sqrt{\half-\abs{a}^2}\right)\left(e^{-i\theta}\ket{ee}-\ket{eg}-\ket{ge}+e^{i\theta}\ket{gg}\right)\right].
\end{eqnarray}

As stated in Sec.~\ref{sec:2qbasin} this state shows no entanglement
between the qubits and the field and therefore the entropy of the
reduced qubit density matrix will be zero. The value of the tangle
for the qubit state at $t=\sfrac{t_r}{2}$ is found to be exactly the
same as the tangle for the initial state of the qubits, therefore
$\tau(0)=\tau(\sfrac{t_r}{2})$. We have called this phenomenon
`Collapse and Revival of Entanglement' \cite{Jarvis08a}. Analogous
to the single qubit case, the quantum information in the initial
state of the qubits (for states within the basin of attraction) is
swapped out and encoded into the field state at the `attractor' times.
Calculations demonstrate that the value of the tangle remains near
zero for long periods between revival. Therefore we are dealing with
a phenomenon which has been described as the `death of entanglement'
by Yu and Eberley \cite{Yu04,Yu09} which forms the centre of much
current interest \cite{Eberly07}.  Qing \emph{et al.} \cite{Qing}
have found a similar collapse and revival for the same model we have
studied here but for very different initial conditions. Even just
within the system we consider here, detailed studies reveal
interesting links between the dynamics of the disappearance (and
reappearance) of entanglement and the basin of attraction, as we
discuss in this and the subsequent subsection.

So if the qubit system starts in the basin of attraction and
$\tau\neq0$, over time the initial entanglement in the qubits is
exchanged for entanglement between the qubits and the field,
and then for a superposition of field states. At the time of the
`attractor' state, $\sfrac{t_r}{4}$, there is absolutely no
entanglement present in the system. Then the process starts to
reverse. Firstly there is again entanglement between the qubits and
the field and at the two qubit revival time, $\sfrac{t_r}{2}$,
$\tau$ returns to its initial value and only qubit entanglement is
present.

\begin{figure}[tp!]
\centering{
\epsfig{file=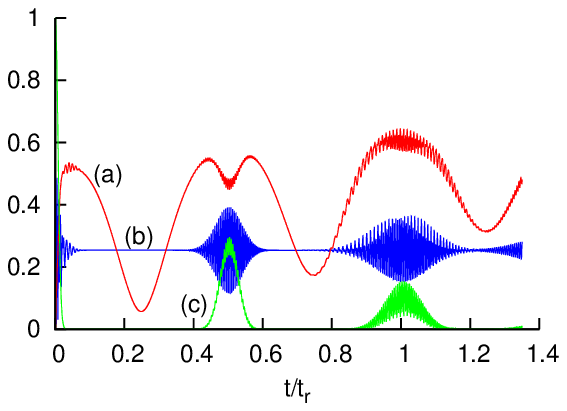,height=8cm}\\
\epsfig{file=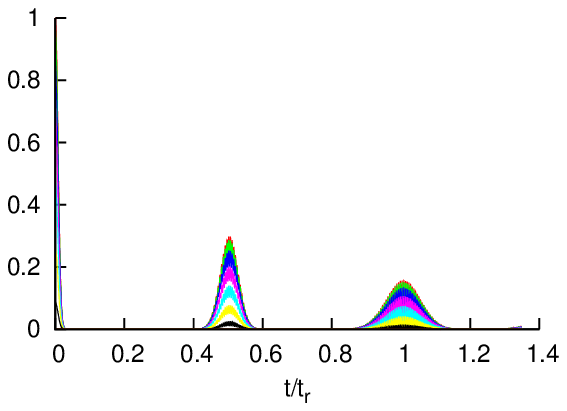,height=8cm}
} \caption{(color online) (Top) `Collapse and Revival of
Entanglement' when the qubits are started in the maximally entangled
state $(\ket{ee}+\ket{gg})/\sqrt{2}$ where $\nbar=50$. (a) the
entropy of the qubits. (b) the probability of the two qubit state $\ket{gg}$. (c) the tangle for the system.\\
(Bottom)`Collapse and Revival of Entanglement' for many different
initial qubit states in the basin of attraction with $\nbar=50$.
These are shown in many different colours depending on the initial
entanglement measure. It shows that the entanglement in the qubits
is initially lost then returns at the first revival peak,
$t=\sfrac{t_r}{2}$. As the initial state is in the basin of
attraction we know the whole system has zero entanglement at
$t=\sfrac{t_r}{4}$. So there is even no entanglement between the
field and the qubits. Therefore we know that over time the amount of
entanglement varies a great deal.\label{fig:reventanglement}}
\end{figure}

The actual extent of the entanglement revival for the exact solution
can be seen in Fig.~\ref{fig:reventanglement}~(a) where we show the
tangle for different initial states within the basin of attraction.
On first glance we see that the amount of tangle present at
$t=\sfrac{t_r}{2}$ is nowhere near the initial amount of tangle
present at $t=0$, which is far from the prediction in the large
$\nbar$ approximation. The value of the entropy at $\sfrac{t_r}{2}$
is around $0.5$ for the case when $a=\sfrac{1}{\sqrt{2}}$, rather
than the predicted value of zero,
Fig.~\ref{fig:reventanglement}~(b). In order for there to be maximum
possible entanglement in the system there has to be the least amount
of entropy, so there is a trade off in tangle for entropy
\cite{Munro}.

Another very interesting feature illustrated in
Fig.~\ref{fig:reventanglement}~(a) is the {\it manner} in which the
entanglement vanishes. For all states within the basin of
attraction, although the initial entanglement decays rapidly (before
reviving at $\sfrac{t_r}{2}$ and later times) it does so with a
Gaussian envelope (as per the Rabi oscillation collapse). It
therefore goes smoothly to zero---this is reflected by the further
observation that the eigenvalue expression in
Eq.~(\ref{eq:mixedtangle}) never goes negative, so the
$\emph{max}$ operation never needs to be implemented for the
tangle. This can be explained by considering the rank of the matrix
used to calculate the tangle,
$\rho_{q}\left(\sigma^y\otimes\sigma^y\right)\rho_{q}^{\ast}\left(\sigma^y\otimes\sigma^y\right)$.
The rank of a matrix is equal to the number of non-zero eigenvalues.
The matrix used to calculate the tangle for this system has a
maximum of rank$=2$ when the qubits are started in the basin of
attraction, so $\lambda_3=\lambda_4=0$. Due to the definition of the
tangle, $\lambda_1\geq\lambda_2$ and Eq.~(\ref{eq:mixedtangle})
becomes:
\begin{equation}
\tau=(\lambda_1-\lambda_2)^2\geq 0.
\end{equation}

This characteristic `Collapse and Revival' behaviour is to be
contrasted with behaviour where entanglement vanishes or appears
with finite gradient and where the $\mathrm{max}$ operation is
needed in Eq.~(\ref{eq:mixedtangle}) because the eigenvalue
expression does go negative. To distinguish these behaviours, we use
the terminology `sudden death/birth' of entanglement \cite{Yu04} for
the latter. This phenomenon does not occur for initial qubit states
within the basin of attraction (which all show collapse and revival
of entanglement) but can for those outside, as will be demonstrated
in the next subsection.

\subsection{Outside the Basin of Attraction}

The previous subsections discussed interesting features of the
entanglement dynamics for the two-qubit JCM when considering initial
conditions inside the basin of attraction. In this subsection we
highlight interesting features that emerge when considering all
other cases---when the qubit Hilbert space is not restricted and,
for example, the terms
$(b\ket{eg}+\sqrt{1-\abs{b}^2}\ket{ge})/\sqrt{2}$, where $b$ is a
complex number, are present in the initial state, as well as
unequal superpositions of $\ket{gg}$ and $\ket{ee}$. To date, in
other works it has been found that the JCM for two qubits can
exhibit an interesting phenomenon called `Sudden Death of
Entanglement', for certain initial conditions of the qubits. Some
investigations start with the qubits and the field in a mixed state
and some others do not \cite{Qing}.

In our work we consider the case when the radiation field is started
in a coherent state. Qubits states inside the basin of attraction
demonstrate collapse and revival of entanglement, but not sudden
death or birth. However, if we venture outside the basin of
attraction and include other initial conditions of the two qubit
system, we find that there is indeed sudden death of entanglement.
This means that we do not restrict the condition $\beta_0=0$ in
Eq.~(\ref{eq:statecfinal}).

\begin{figure}[tp]\center{
\epsfig{file=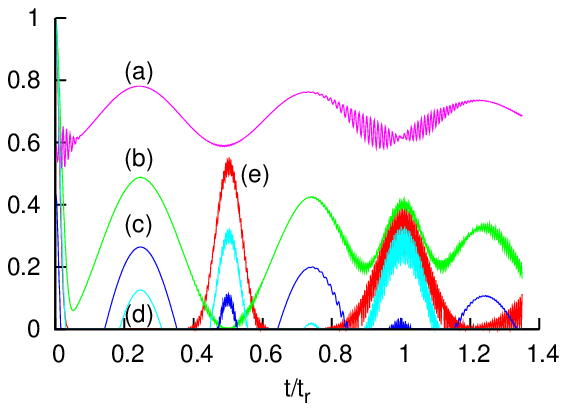, height=8cm}\\
\epsfig{file=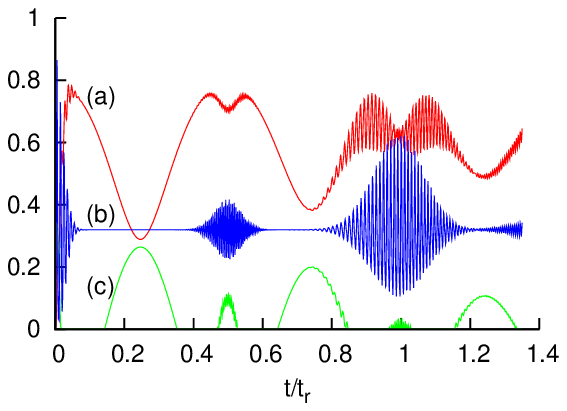, height=8cm}
} \caption{(color online) (Top) Values of the concurrence over time
for several different initial qubit states. As in previous figures,
$\bar n = 50$.  (a) The initial state of the qubits is
$\frac{1}{\sqrt{10}}\ket{ee}-\sqrt{\frac{9}{10}}\ket{gg}$. At times
the concurrence is greater than the initial value. (b) The initial
state of the qubits is
$\frac{1}{\sqrt{2}}\left(\ket{eg}+i\ket{ge}\right)$, and
unlike initial conditions within the basin of attraction, leads
to a concurrence value of zero at $\sfrac{t_r}{2}$. (c) The initial
state of the qubits is
$\frac{1}{\sqrt{20}}\ket{ee}+\sqrt{\frac{19}{20}}\ket{gg}$ and shows
`Sudden Death of Entanglement'. (d) The initial state of the qubits
is $\frac{1}{\sqrt{2}}\left(\ket{eg}+e^{\frac{i
\pi}{4}}\ket{ge}\right)$, a maximally entangled state that exhibits
`Sudden Death of Entanglement'.
(e) The initial state of the qubits is $\frac{1}{\sqrt{2}}\left(\ket{ee}+\ket{gg}\right)$, a maximally entangled state that lies in the basin of attraction that is used as a comparison. \\
(Bottom) `Sudden Death of Entanglement' when the qubits are started
in the state
$\frac{1}{\sqrt{20}}\ket{ee}+\sqrt{\frac{19}{20}}\ket{gg}$, case (c)
in the top diagram. As before (a) the entropy of the qubits. (b) the
probability of the qubit being in the state $\ket{gg}$. (c) the
mixed state concurrence of the qubit system.
\label{fig:synnotbasin}}
\end{figure}

Now, without this restriction, the qubit system will not go towards
the `attractor' state at $\sfrac{t_r}{4}$ and $\frac{3t_r}{4}$ in the
large $\nbar$ approximation as $\beta_0\neq0$. Furthermore the
entropy will be non zero at these times as the qubit state can not
be factorised out of the wavefunction. However, the system
will still have zero entropy at the `full' revival time $t_r$
as
$\ket{\Phi_{-1}(t_r)}=\ket{\Phi_{0}(t_r)}=\ket{\Phi_{+1}(t_r)}=\ket{\alpha}$
and the field state can be factorised out of the wavefunction. Like
before $\tau(0)=\tau(t_r)$ so the amount of entanglement between the
qubits is the same at the main revival time and $t=0$. For
$\nbar=50$, this revival will not be complete and so the tangle
will not return to its initial value. Fig.~\ref{fig:synnotbasin}
shows examples of the qubit entanglement dynamics for initial states
which are outside the basin of attraction for the numerically evaluated exact solution. In this figure a new measure of qubit entanglement has
been used called the concurrence $\zeta=\sqrt{\tau}$ and has been
used as a direct comparison to previous work
\cite{Eberly07,Yu09,Yonac}.

Remarkably, Fig. \ref{fig:synnotbasin} shows that a field initially in a coherent
state can give sudden death and birth of entanglement, with initial
qubit states of various forms lying outside the basin of attraction.
A case where the initial conditions are in the basin of attraction
is shown for reference. Clearly peaks in the entanglement measure at
both $\sfrac{t_r}{4}$ and $\frac{3t_r}{4}$ arise (with death/birth
at finite gradient)---times when there would be no entanglement if
the initial qubit state was in the basin of attraction. There are
still peaks at the times $\sfrac{t_r}{2}$ and $t_r$, but these have
a lower maximum and also now exhibit birth/death behaviour rather
than Gaussian collapse and revival. One initial condition that
yields `Sudden Death of Entanglement' has been shown in
Fig.~\ref{fig:synnotbasin} alongside an entropy measurement.

To recap, we find that our system can display both a sudden
birth/death of entanglement as well as a Gaussian collapse and
revival of entanglement, depending on initial conditions.
Furthermore, we understand that, in this system, the sudden death
occurs when $\lambda_1-\lambda_2-\lambda_3-\lambda_4$ becomes
negative and so, according due to the \emph{max} operation used in
Eq.~\ref{eq:mixedtangle} the tangle goes to zero with a finite
gradient. In contrast, within the basin of attraction, there are
only two non-zero eigenvalues $\lambda_1>\lambda_2$, so the tangle
can only display a smooth collapse.

These additional peaks in the qubit entanglement seen in
Fig.~\ref{fig:synnotbasin} are due to the inclusion of $\ket{D_0}$
and $\ket{\Phi_0(t)}$ in the wavefunction. From the analysis in Sec.
\ref{sec:2qubit} we showed that $\ket{D_{\pm
1}(\sfrac{t_r}{4})}=\twoattp$, the unentangled two qubit `attractor'
state, so all entanglement present at $t=\sfrac{t_r}{4}$ is due to
$\ket{D_0}$. If $\ket{D_0}$ is a product state then there will be no
entanglement in the qubit system. However if $\ket{D_0}$ is
entangled it still does not mean there will be some qubit
entanglement as a mixture of entangled and product states need not
be entangled. For example curve (b) in the top diagram of
Fig.~\ref{fig:synnotbasin} shows there is zero concurrence at
$\frac{t_r}{2}$, even though $\ket{D_0}$ contains some entanglement.

\section{$N_q$ qubit Jaynes-Cummings Model}\label{sec:nqqubits}

Having found a sense in which the one qubit `attractor' state has
analogues for two qubits, it is natural to inquire whether similar
generalisation holds for the case of more than two qubits. We use
the Hamiltonian shown in Eq. (\ref{eq:Nqham}) for $N_q$ qubits. We
have found that there are indeed cases where the system goes to a
state similar to the one qubit `attractor' state and have a general
form for the basin of attraction for $N_q$ qubits.

In the large $\nbar$ approximation the state of the system takes the
following form \cite{Meunier}:
\begin{equation}
\ket{\widetilde{\Psi}_{N_q}(t)}=\sum_{k=-N_q/2}^{N_q/2}\beta_k(t)\ket{D_k(t)}\otimes\ket{\Phi_k(t)}.
\end{equation}
$\ket{D_k(t)}$ has a complicated form and occurs in the paper by
Meunier \emph{et al.} \cite{Meunier}. For this work we will be
most concerned with the states that appear in the expression for the
basin of attraction. These take the following simple form:
\begin{eqnarray}
\ket{D_{\pm\frac{N_q}{2}}(t)}&=&\ket{D_{\pm\half}(N_q\, t)}\otimes \ldots\otimes\ket{D_{\pm\half}(N_q\, t)}\\
&=&\ket{D_{\pm\half}(N_q\, t)}^{\otimes N_q}.\label{eq:Dnq}
\end{eqnarray}

Note that:
\begin{equation}\label{eq:phinq}
\ket{\Phi_k(t)}=\ket{e^{i\pi k N_q t/t_r}\alpha},
\end{equation}
and we can use this equation to understand the simple dynamics of
the $N_q$ solution. The field initially starts in a coherent state
that then splits into $N_q+1$ different `blobs' and a revival peak
occurs every time at least two of these wavepackets overlaps. The
strength of the revival is determined by the number of wavepackets
that overlap and these revival peaks occur at definite intervals.
Using Eq. (\ref{eq:phinq}) we can write a general form for the
revival times:
\begin{equation}\label{eq:revivaltimes}
\left(k+\frac{1}{p}\right)t_r,\ \ \
\left(k+\frac{(p-1)}{p}\right)t_r,\ \ \ p=1,2,..., N_q \ \
{\rm{and}} \ \  k=1,2,..., \infty.
\end{equation}

Eq. (\ref{eq:phinq}) also highlights that when there are an even
number of qubits the field returns to its initial state
$\ket{\alpha}$ after a time of $t_r$. The time till the wavepackets
return to their original position is determined by the slowest
moving wavepackets, which are $\ket{e^{\pm 2 i\pi t/t_r}\alpha}$ in
the even qubit case and $\ket{e^{\pm i\pi t/t_r}\alpha}$ in the odd
qubit case. When $N_q$ is odd the wavepackets return to their start
position after a time $2t_r$ and there will always be no stationary
wavepacket at the start position.

\begin{figure}[tp]
\centering{
\epsfig{file=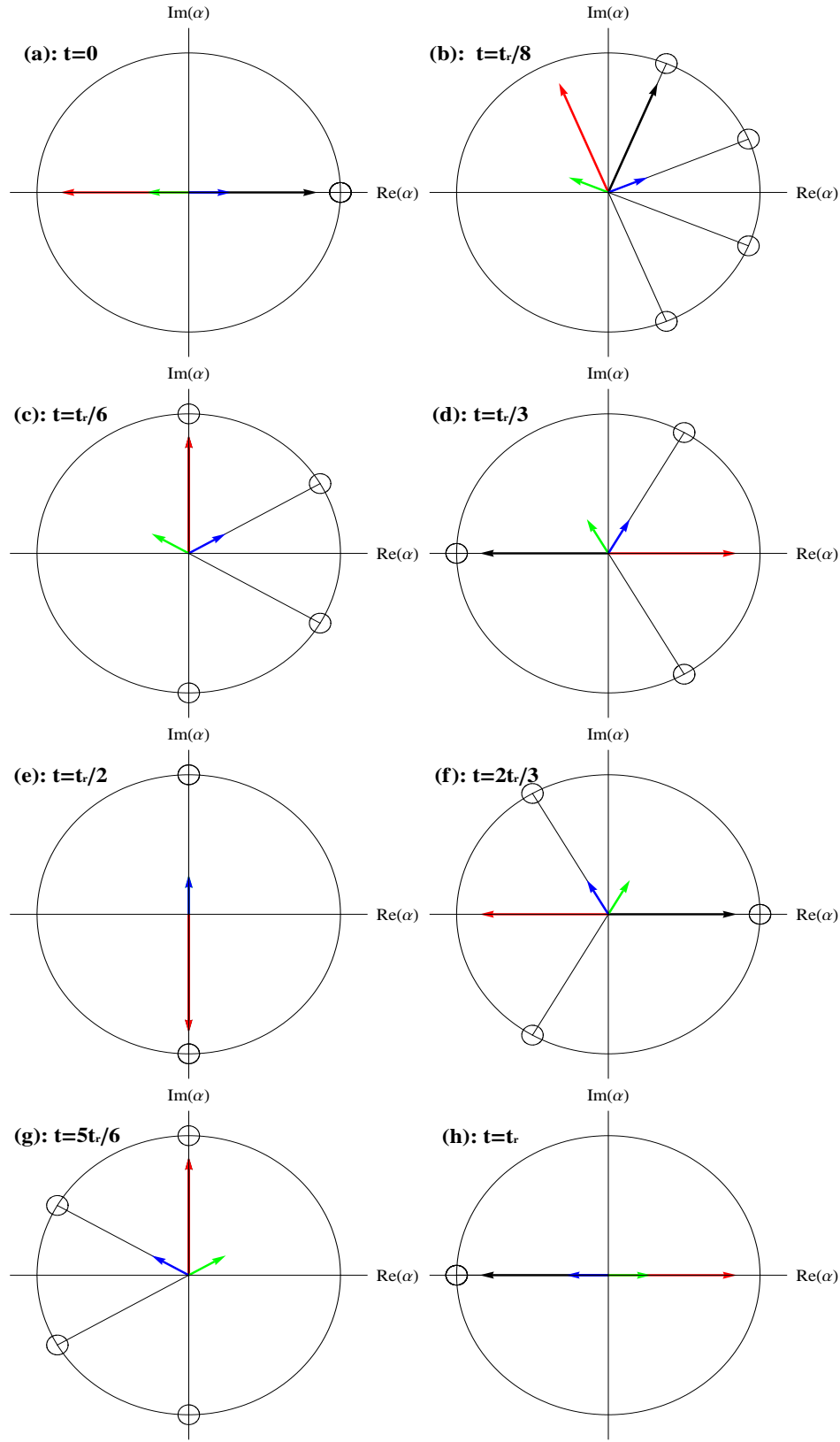,height=22cm}
\caption{(color online) Phase space sketches of the $Q$ function at
eight different times when the qubits are initial started in the
state $\ket{ggg}$. The atomic dipole states are represented as
arrows. (a) the time $t=0$. The system is started in a coherent
state which is shown by a circle of uncertainty in phase space. (b)
a time a little after $t=0$. (c) the time $t=\sfrac{t_r}{6}$. (d)
the time $t=\sfrac{t_r}{3}$. (e) the time $t=\sfrac{t_r}{2}$. (f)
the time $t=\sfrac{2t_r}{3}$. (g) the time $t=\sfrac{5t_r}{6}$. (h)
the time $t=t_r$. } \label{fig:3qdipole} }
\end{figure}

A sketch of the $Q$ function of the field for three qubits is shown
in Fig. \ref{fig:3qdipole} for large values of $\nbar$ and we once
again show the dipole moments as arrows. Using $\ket{D_k}$ in
reference \cite{Meunier} we can compute a general term for the
equation of the dipole moments in the large $\nbar$ approximation
and is shown below.
\begin{eqnarray}
\overrightarrow{d}_{k}&=&\bra{D_{k}}\overrightarrow{\hat{J}}_{N_{q}}\ket{D_k}=\bra{D_{k}}\overrightarrow{\hat{\sigma}}_{N_{q}}\ket{D_k}\nonumber\\
&=&\abs{k}\left(\pm \cos\left(\theta+\frac{2\abs{k}\pi
t}{t_r}\right) \underline{\hat{i}}+
\sin\left(\theta+\frac{2\abs{k}\pi
t}{t_r}\right)\underline{\hat{j}}\right)
\end{eqnarray}
where
$\overrightarrow{\hat{\sigma}}_{N_{q}}=\hat{\sigma}_{N_{q}}^x\underline{\hat{i}}+\hat{\sigma}_{N_{q}}^y\underline{\hat{j}}+\hat{\sigma}_{N_{q}}^z\underline{\hat{k}}$,
is the Pauli vector for $N_{q}$ qubits and
\begin{eqnarray}
\hat{\sigma}_{N_{q}}^{x,y,z}&=&\sum^{N_{q}}_{i=1}\hat{\sigma}^{x,y,z}_i.
\end{eqnarray}

We notice again the pattern of the arrows changes over time and that
the wavepackets come to their original starting position after a
time $2t_r$.

\subsection{Basin of Attraction}\label{sec:nqbasin}

Using the above experience with one and two qubits, we investigate
whether a basin of attraction exists for $N_q>2$. Using the
expansion pioneered by Chumakov \emph{et al.} \cite{Chumakov94} and
developed independently by Meunier \emph{et al.} \cite{Meunier} we
have indeed found a basin of attraction for $N_q$ qubits. As
$\ket{D_{\pm \half}(\sfrac{t_r}{2})}=\attractp$, and using the
observation made in Eq. (\ref{eq:Dnq}), we can therefore conclude
that $\ket{D_{\pm \frac{N_q}{2}}(t)}$ will go to a direct product of
$\attractp$ at $t=\sfrac{t_r}{2 N_q}$.

\begin{eqnarray}
\ket{D_{\pm\frac{N_q}{2}}\left(\sfrac{t_r}{2 N_q}\right)}&=&\ket{D_{\pm\half}(\sfrac{t_r}{2})}^{\otimes N_q}\\
&=&\ket{\psi^{+}_{1\,att}}^{\otimes N_q}\\
&=&\ket{\psi^{+}_{N_q\,att}}\label{eq:nqattract}
\end{eqnarray}
By setting the values of $\beta_p(t)=0$, for
$p=\sfrac{N_q-2}{2},...,-\sfrac{N_q-2}{2}$ then only $\ket{D_{\pm
\frac{N_q}{2}}(t)}$ are in the wavefunction and the qubits will go
to $\ket{\psi_{N_q\,att}^+}$ at $t=\sfrac{t_r}{2 N_q}$. Using this
information we have found a basin of attraction for $N_q$ qubits.

\begin{eqnarray}
{\left\vert \psi _{N_{q}}\right\rangle } &=&\sum_{m=-N_q/2}^{N_{q}/2}\frac{%
A(N_{q},a,m)e^{-i(\frac{N_{q}}{2}-m)\theta }\sqrt{N_{q}!}}{\sqrt{\left(\frac{N_q}{2}+m\right)!\left(\frac{N_q}{2}-m\right)!}}%
{\left\vert N_{q},m\right\rangle }  \nonumber
\label{eq:Nqbasin} \\
A(N_{q},a,m) &=&\left\{
\begin{array}{cc}
a & \text{if $\left(\frac{N_q}{2}-m\right)$ is even} \\
&  \\
\sqrt{\frac{1}{2^{N_{q}-1}}-{\left\vert a\right\vert }^{2}} &
\text{if $\left(\frac{N_q}{2}-m\right)$
is odd}%
\end{array}%
\right.
\end{eqnarray}%
As before $\phi$ is an arbitrary phase and $\theta$ is the initial
phase of the radiation field with
$0\leq\abs{a}\leq\frac{1}{\sqrt{2^{N_q-1}}}$. The basin of
attraction is in the symmetric subspace and $\ket{N_q,m}$ is the
symmetrised state with $N_e$ excited qubits, $N_g$ qubits in the
ground state and $m=\frac{N_e-N_g}{2}$. For example, for the three
qubit case,
$\ket{N_q=3,m=\half}=\sfrac{1}{\sqrt{3}}\left(\ket{eeg}+\ket{ege}+\ket{gee}\right)$
as $N_e=2$ and $N_g=1$ etc. We note that although the state
space of the qubits increases with increasing $N_q$, the basin of
attraction is still parameterised by the single complex value $a$.
The significance of this will become apparent in Sec.
\ref{sec:catstates}.

We emphasise that when the qubits start in the basin of
attraction then there are only two wavepackets in phase space of
the photon field for all time. The initial conditions have
restricted the system to behave like the one qubit case as is
highlighted by considering the wavefunction at
$t=\sfrac{t_r}{2N_q}$.
\begin{equation}
\ket{\widetilde{\Psi}_{N_q}\left(\sfrac{t_r}{2N_q}\right)}=\ket{\psi^+_{N_q\,att}}\left(\beta_{\sfrac{N_q}{2}}\left(\sfrac{t_r}{2N_q}\right)\ket{i\alpha}+\beta_{-\sfrac{N_q}{2}}\left(\sfrac{t_r}{2N_q}\right)\ket{-i\alpha}\right)
\end{equation}
In the one qubit case the qubit system also goes to a second
`attractor' state $\attractm$ at the time $\sfrac{3t_r}{2}$ and
so $\ket{D_{\pm \frac{N_q}{2}}\left(\frac{3t_r}{2
N_q}\right)}=\ket{\psi^{-}_{1\,att}}^{\otimes
N_q}=\ket{\psi^{-}_{N_q\,att}}$. Qubits started in the basin of
attraction, Eq. (\ref{eq:Nqbasin}), will go to these `attractor'
states and hence the entropy will go to zero.

We can construct a series for when these `attractor' times will
occur for $N_q$ qubits, shown in table \ref{table:revandent} with
the revival times from Eq. (\ref{eq:revivaltimes}). The general
equation for these times is:
\begin{equation}
\left(k+\frac{(2p-1)}{2N_q}\right)t_r,\ \ \ \ \ \ p=1,2,..., N_q \ \
{\rm{and}} \ \  k=1,2,..., \infty.
\end{equation}

\begin{table}
\centering
\begin{tabular}{|c|c|c|c|}
\hline
No. of qubits&No. of Blobs&Revival Times&Possible `Attractor' Times\\
\hline
1&2& $t_r$&$\frac{t_r}{2}$\\
2&3&$\frac{t_r}{2}$, $t_r$ &$\frac{t_r}{4}$, $\frac{3t_r}{4}$\\
3&4&$\frac{t_r}{3}$, $\frac{t_r}{2}$, $\frac{2t_r}{3}$, $t_r$&$\frac{t_r}{6}$, $\frac{t_r}{2}$, $\frac{5t_r}{6}$\\
4&5&$\frac{t_r}{4}$, $\frac{t_r}{3}$, $\frac{t_r}{2}$, $\frac{2t_r}{3}$, $\frac{3t_r}{4}$, $t_r$&$\frac{t_r}{8}$, $\frac{3_r}{8}$, $\frac{5t_r}{8}$, $\frac{7t_r}{8}$\\
5&6&$\frac{t_r}{5}$, $\frac{t_r}{4}$, $\frac{t_r}{3}$, $\frac{t_r}{2}$, $\frac{2t_r}{3}$, $\frac{3t_r}{4}$, $\frac{4t_r}{5}$, $t_r$&$\frac{t_r}{10}$, $\frac{3t_r}{10}$, $\frac{t_r}{2}$, $\frac{7t_r}{10}$, $\frac{9t_r}{10}$\\
\hline
\end{tabular}
\caption{A summary of where the different revival peaks occur for
differing numbers of qubits. Also the times are shown for when
$\ket{D_{\pm\frac{N_q}{2}}\left(t\right)}=\ket{\psi^{\pm}_{N_q\,att}}$.
Note that only revival and `attractor' times in the interval
$0<t\leq t_r$ are shown. Further times can be obtained by adding
integer multiples of $t_r$. \label{table:revandent}}
\end{table}

\begin{figure}[tp]\center{
\epsfig{file=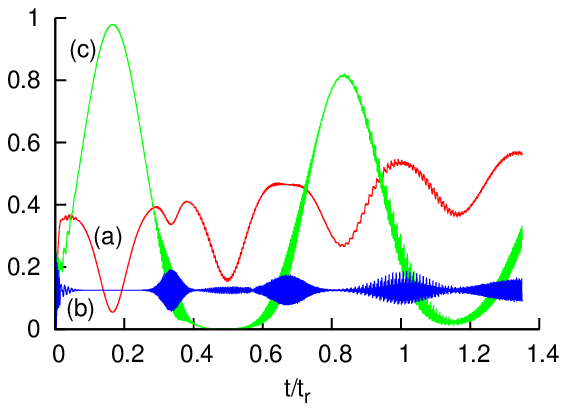, height=8cm}\\
\epsfig{file=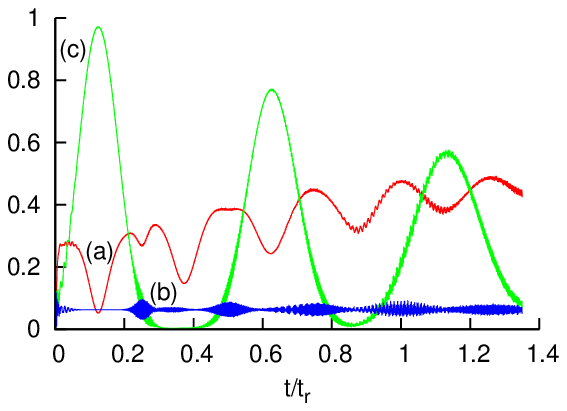, height=8cm}
} \caption{(color online) Time evolution of the system when the
initial qubit state is in the basin of attraction for $\nbar=50$ and
$\theta=0$.
 (Top) Three qubits: The initial state of the qubits is $\half\left(\ket{N_q=3,m=\sfrac{3}{2}}+\sqrt{3}\ket{N_q=3,m=-\half}\right)$.
(Bottom) Four qubits: The initial state of the qubits is
$\frac{1}{\sqrt{2}}(\ket{N_q=4,m=1}+\ket{N_q=4,m=-1})$. As before
(a) the entropy of the qubits. (b) the probability of all the qubits
being in the state $\ket{g}$. (c) the probability of being in the
qubit `attractor' state $\nqattp$. \label{fig:3&4qub}}
\end{figure}

In Fig. \ref{fig:3&4qub} we show the $3$ and $4$ qubit evolution
starting in a state in the basin of attraction (Eq.
(\ref{eq:Nqbasin})). In general we do see there are dips in the
entropy that go close to zero and the probability of being in the
`attractor' state goes to close to one. These features get more
pronounced as $\nbar$ is increased. As for the two-qubit case, when
this probability goes to zero, the system is in the second
`attractor' state, $\ket{\psi^{-}_{N_q\,att}}$, which is orthogonal
to the first.

\subsection{Collapse and Revival of `Schr\"{o}dinger Cat' States}\label{sec:catstates}

Surprisingly, the basin of attraction has a simple form when written
in terms of spin coherent states, which are analogous to
coherent states of a photon field. As described in Appendix
\ref{ap:spin}, such states are given by:
\begin{eqnarray}
\ket{z,N_q}&=&\frac{1}{(1+\abs{z}^2)^{N_q/2}}\sum_{m=-N_q/2}^{N_q/2}\sqrt{\frac{N_q!}{(\frac{N_q}{2}-m)!(\frac{N_q}{2}+m)!}}z^{(\frac{N_q}{2}-m)}\ket{N_q,m}\\
&=&\frac{1}{(1+\abs{z}^2)^{N_q/2}}\left(\ket{e}+z\ket{g}\right)^{\otimes
N_q}\label{eq:majorana}
\end{eqnarray}
where $\ket{N_q,m}$ is defined in Sec. \ref{sec:nqbasin}. A
spin coherent state is a product state and this is easily seen when
it is written in Majorana representation \cite{Majorana32} in Eq.
(\ref{eq:majorana}). Below we rewrite Eq. (\ref{eq:Nqbasin}) in the
form of a superposition of two spin coherent states, and show
the form of the wavefunction at the time $t=0$ when the qubits are
in the basin of attraction:
\begin{eqnarray}\label{eq:nqbasinspin}
\ket{\widetilde{\Psi}_{N_q}(0)} =\ket{\psi_{N_q}}\ket{\alpha}&=&\left(\sqrt{2^{N_q-2}}\left(a+\sqrt{\frac{1}{2^{N_{q}-1}}- {\abs{a}}^{2}}\right)\ket{z=e^{-i\theta},N_q}\right.\nonumber\\
&+&\left.\sqrt{2^{N_q-2}}\left(a-\sqrt{\frac{1}{2^{N_{q}-1}}-
{\abs{a}}^{2}}\right)\ket{z=-e^{-i\theta},N_q}\right)\ket{\alpha}
\end{eqnarray}

From Eq. (\ref{eq:majorana}) it is clear that
$\ket{z=e^{-i\theta},N_q}$ is orthogonal to
$\ket{z=-e^{-i\theta},N_q}$ and $\ket{\psi_{N_q}}$ is a collection
of spin `Schr\"{o}dinger cat' states. We also write the state of the
field at the time of the first dip in the entropy for an arbitrary
number of qubits, $t=\sfrac{t_r}{2N_q}$:
\begin{eqnarray}\label{eq:nqattrattime}
\ket{\widetilde{\Psi}_{N_q}(\sfrac{t_r}{2N_q})} &=&\nqattp\left(\sqrt{2^{N_q-2}}\left(a-\sqrt{\frac{1}{2^{N_{q}-1}}- {\abs{a}}^{2}}\right)e^{i \pi\nbar/2}\ket{i\alpha}\right.\nonumber\\
&&-\left.\sqrt{2^{N_q-2}}\left(a+\sqrt{\frac{1}{2^{N_{q}-1}}-
{\abs{a}}^{2}}\right)e^{-i \pi\nbar/2}\ket{-i\alpha}\right)
\end{eqnarray}

Eq. (\ref{eq:nqbasinspin}) and Eq. (\ref{eq:nqattrattime}) are very
similar and show how a Schr\"{o}dinger cat state, and the
information about the initial conditions of the qubits, moves from
the qubits to the field. At the time $t=0$ the qubits are in a spin
`Schr\"{o}dinger cat' state and the variable $a$, which fixes the
initial state of the qubits, determines the ratio of
$\ket{z=e^{-i\theta},N_q}$ to $\ket{z=-e^{-i\theta},N_q}$. At the
time $t=\sfrac{t_r}{2N_q}$ the field is now in a `Schr\"{o}dinger
cat' state and the information about the initial state of the qubit
encoded by $a$ is now stored in the field state. A `Schr\"{o}dinger
cat' state, and the information about the initial qubit state has
moved from the qubit system to the field system \cite{Jarvis08a}.

\begin{figure}
\centering{
\epsfig{file=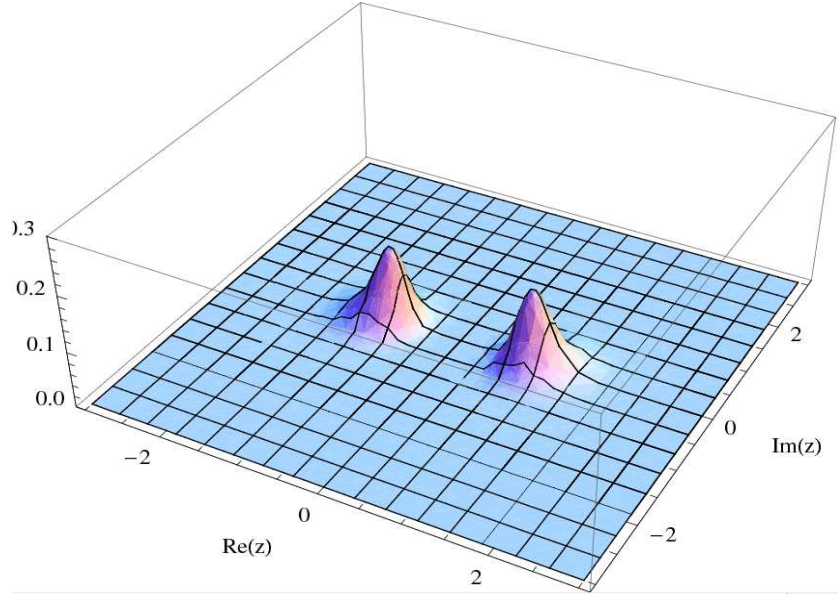,height=8cm}
\caption{(color online) A plot of the spin $Q_q$ function of the
qubits at the time $t=0$ for a state in the basin of
attraction. The value of $a=0$ and we see a perfect Schr\"{o}dinger
cat state of the qubits. The number of qubits for the plot is
$N_q=40$ with $\theta=0$, but a similar picture can be produced for
any $N_q>3$. The peaks merge together for $N_q\leq3$.
\label{fig:catofqubits} }}
\end{figure}

To illustrate this migration we use a function similar to the field
$Q$ function, $Q_q$, using the completeness relation in appendix
\ref{ap:spin},
\begin{equation}
Q(\alpha,t)=\frac{\bra{\alpha}\rho_f(t)\ket{\alpha}}{\pi}\ \ \ \ \ \
Q_{q}(z,t)=\frac{\bra{z}\rho_q(t)\ket{z}}{(1+\abs{z}^2)^2}
\end{equation}
$Q_{q}(z,t)$ corresponding to $\ket{\psi_{N_q}}$ in Eq.
(\ref{eq:nqbasinspin}) is depicted in Fig \ref{fig:catofqubits}.
There are indeed two separate peaks that correspond to two
macroscopically different spin coherent states.

The `attractor' states for $N_q$ qubits in Eq. (\ref{eq:Nqbasin})
are also spin coherent states (see appendix \ref{ap:spin}).
\begin{eqnarray}\label{eq:Nqattractb}
\ket{\psi^{\pm}_{N_q\,att}}&=&\frac{1}{\sqrt{2^{N_q}}}\left(e^{-i\theta}\ket{e}\pm i\ket{g}\right)\otimes...\otimes\left(e^{-i\theta}\ket{e}\pm i\ket{g}\right)\nonumber\\
&=&\frac{1}{\sqrt{2^{N_q}}}\left(e^{-i\theta}\ket{e}\pm
i\ket{g}\right)^{\otimes
N_q}\nonumber\\
&=&\ket{z=\pm ie^{i\theta},N_q}
\end{eqnarray}

\begin{figure}
\centering{
\epsfig{file=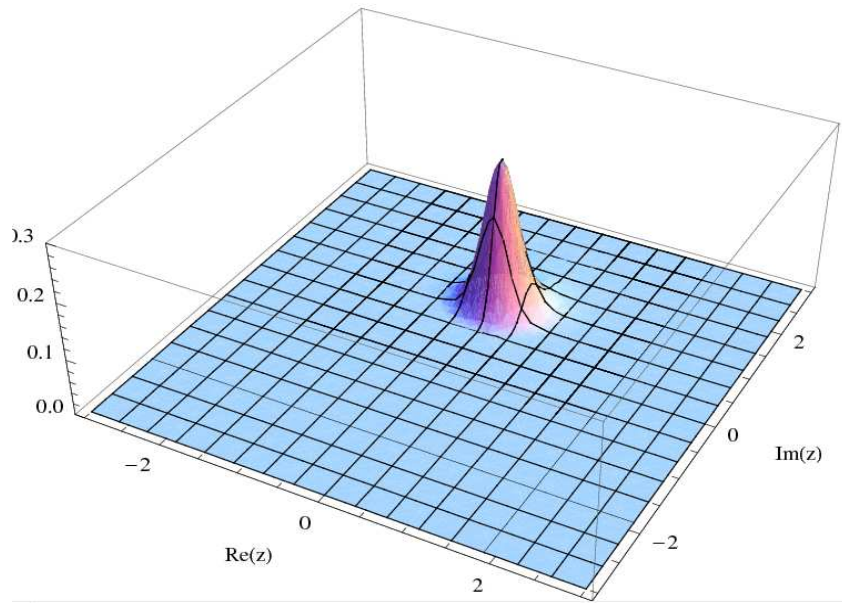,height=8cm}
\caption{(color online) A plot of the spin $Q_q$ function of the
qubits at the time $t=t_r/80$, the `attractor' time when $N_q=40$. The
initial value of $a=0$ and we see a spin coherent state centered
around the value $z=ie^{i\theta}$ with $\theta=0$.
\label{fig:attractqubits} }}
\end{figure}

In Fig \ref{fig:attractqubits} we plot a spin $Q_q$ function diagram
for the `attractor' state.

So far we have shown, for large $\nbar$, that the qubit states in
the basin of attraction are spin Schr\"{o}dinger cat states when
$a\neq\sfrac{1}{\sqrt{2^{N_q}}}$. At a later time,
$t=\sfrac{t_r}{2N_q}$, the qubits are in the spin coherent
`attractor' state and the field is in a Schr\"{o}dinger cat state.
Then the qubits return once again to a spin Schr\"{o}dinger cat
state at the time $t=\sfrac{t_r}{N_q}$. This is shown in Table
\ref{table:cats}.
\begin{table}[ht]
\centering
\begin{tabular}{|c|c|c|c|}
\hline
&$t=0$&$t=t_r/2N_q$&$t=t_r/N_q$\\
\hline
Field&Coherent State& Schr\"{o}dinger cat state&Coherent State\\
Qubit&Spin Schr\"{o}dinger cat state&Spin Coherent State&Spin Schr\"{o}dinger cat state\\
\hline
\end{tabular}
\caption{A table of the states of the field and qubits at different
times. The behavior repeats in the large $\nbar$ approximation.
\label{table:cats}}
\end{table}
So we conclude that the `Schr\"{o}dinger cat' state migrates from
the qubits to the field and back again.
\subsection{Collapse and Revival of Entanglement}
Whilst the basin of attraction for the $N_q$ qubit case is a very
small part of the Hilbert space of all initial states it shows many
different degrees of entanglement, just like the two qubit case.
Some values of $a$ and $\phi$ give the qubit state to be in a
product state and other values give some degree of entanglement
between the qubits. When considering more than two qubits there is
no universally accepted measure of entanglement. Indeed, for
$N_q\geq4$ and multilevel systems there exist infinitely many
inequivalent kinds of entanglement \cite{Plenio}. GHZ type
entanglement exists for all numbers of qubits and, up to local
unitaries, a maximally entangled GHZ state has the form:
\begin{equation}
\ket{GHZ}=\sfrac{1}{\sqrt{2}}\left(\ket{b}^{\otimes
N_q}+\ket{b_{\bot}}^{\otimes N_q}\right)
\end{equation}
where $\braket{b}{b_{\bot}}=0$. In the previous section it was
stated that the basin of attraction can be rewritten as a spin
`Schr\"{o}dinger cat' state which is a state that is a superposition
of two macroscopically distinguishable states. However expanding
states in the basin of attraction using Eq. (\ref{eq:majorana}) we
see that it has a GHZ form:
\begin{eqnarray}\label{eq:GHZbasin}
\ket{\psi_{N_q}} &=&\frac{1}{2}\left(\left(a+\sqrt{\frac{1}{2^{N_{q}-1}}- {\abs{a}}^{2}}\right)\left(\ket{e}+e^{-i\theta}\ket{g}\right)^{\otimes N_q}+\left(a-\sqrt{\frac{1}{2^{N_{q}-1}}- {\abs{a}}^{2}}\right)\left(\ket{e}-e^{-i\theta}\ket{g}\right)^{\otimes N_q}\right)\nonumber\\
\end{eqnarray}
although it is only a maximal GHZ state when $a=0$ and
$a=\sfrac{1}{\sqrt{2^{N_q-1}}}$. Excluding the case when
$a=\sfrac{1}{2^{N_q-1}}$, all other values of $a$ give an entangled
state of GHZ form. So, as with the previous case, there are only two
values of $a$ that give a product state when in the basin of
attraction, meaning all other states contain some entanglement. We
next consider the case of three qubits in more detail and verify by
numerical calculations that the basin of attraction for three qubits
is indeed only made up of GHZ type entanglement.

For three qubits there are two different types of entanglement, GHZ
\cite{GHZ} and W \cite{Dur}.
$\ket{GHZ}=\frac{1}{\sqrt{2}}\left(\ket{eee}+\ket{ggg}\right)$ can
be regarded as a maximally entangled state, but if one of the qubits
is traced out the remaining state has absolutely no entanglement
present. The entanglement of
$\ket{W}=\frac{1}{\sqrt{3}}\left(\ket{eeg}+\ket{ege}+\ket{gee}\right)$
is maximally robust and is not so fragile to particle loss
(i.e. there is still entanglement present after tracing out any
individual qubit). It is not possible to transform $\ket{GHZ}$ into
$\ket{W}$ even with a small probability of success and vise versa,
and thus the two states represent fundamentally different
\emph{types} of entanglement. In particular, the GHZ state
represents a truly 3-body entanglement. Similarly, any $N_q$-qubit
GHZ state contains specifically $N_q$-body entanglement.

\begin{figure}[h!]
\centering{
\epsfig{file=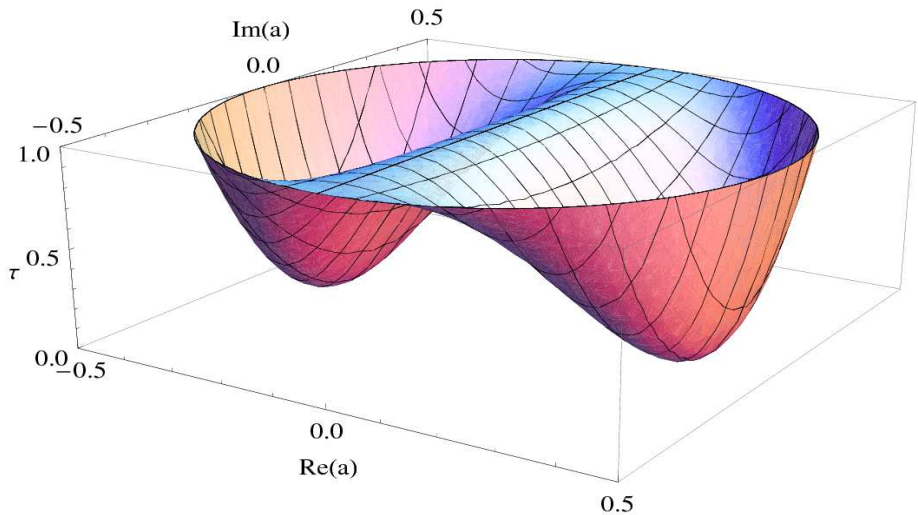,height=7cm}
\caption{(color online) The value of the tangle for the states in
the basin of attraction for different values of $a$. We notice that
the plot shows only two points where the tangle is zero, just like
the two qubit case.} \label{fig:concur3q}}
\end{figure}

We have used a pure state GHZ entanglement measure introduced by
Coffman \emph{et al.} \cite{Coffman}:
\begin{equation}\label{eq:3qpuremeasure}
\tau_{ABC}=C^2_{A(BC)}-C^2_{AC}-C^2_{AB}
\end{equation}
where $\tau_{ABC}$ is the amount of GHZ entanglement,
$C^2_{A(BC)}=2\sqrt{det \rho_A}$ and $C^2_{AC}$ is the entanglement
after qubit B has been traced out. The results for different values
of $a$ for the three qubit version of Eq. (\ref{eq:Nqbasin}) are
shown in Fig. \ref{fig:concur3q}. This measurement only works for
three qubit states that are pure, so we only perform the calculation
on the initial state.

This plot shows the same shape as that of Fig. \ref{fig:concur}, but
values of $a$ are now limited to a smaller range. On the basis of
Figs. \ref{fig:concur} and \ref{fig:concur3q}, and using Eq.
(\ref{eq:GHZbasin}), we conjecture that for more qubits the basin of
attraction will always contain maximally entangled states that
involve all the qubits and two unentangled states, but the size of
the basin will get smaller as the number of qubits increases.

For the large $\nbar$ approximation it was shown in Sec.
\ref{sec:2qcollapseent} that if a state is started in the basin of
attraction $\tau(0)=\tau(\sfrac{t_r}{4})$. This can also be extended
to the three qubit case and
$\tau_{ABC}(0)=\tau_{ABC}(\sfrac{t_r}{6})$. Therefore if the initial
qubit state is in the basin of attraction and $\tau_{ABC}\neq0$ then
the whole entanglement will return to the qubit system at
$t=\sfrac{t_r}{6}$.

\begin{figure}[h]
\centering{
\epsfig{file=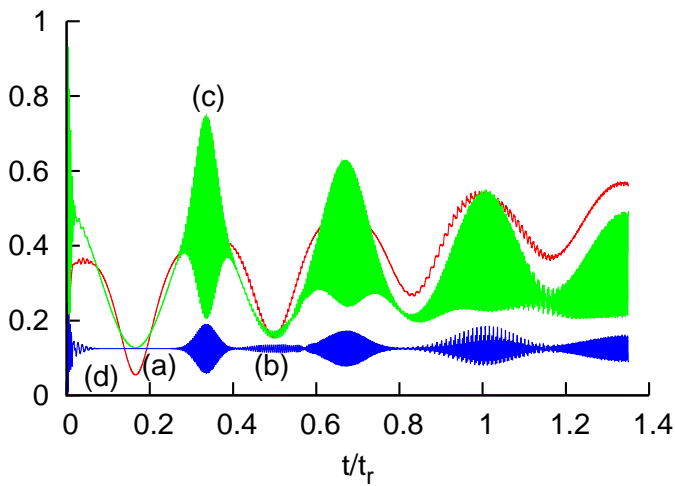,height=7cm}
\caption{(color online) `Collapse and Revival of Entanglement' for
three qubits for the initial state,
$\frac{1}{2}\left(\sqrt{3}\ket{m=1/2}+\ket{m=-3/2}\right)$, which is
inside the basin of attraction, with $\nbar=50$ and $\theta=0$. (a)
the entropy of the qubits. (b) the probability of all the qubits
being in the state $\ket{g}$. (c) the probability of being in the
initial state, $P_{init}(t)$. (d) the value of tangle when one qubit
has been traced out. In fact this pink line remains zero for all
time, so there is no $W$ state entanglement present.
\label{fig:reventangle3}}}
\end{figure}

To see if `Collapse and Revival' of entanglement also occurs in the
exact solution for three qubits we start the qubits in an initial
state, $\ket{\psi_3}$, that is a maximally entangled GHZ state
($\tau_{ABC}=1$). We then calculated the probability
$P_{init}(t)=\bra{\psi_3}\rho_q(t)\ket{\psi_3}$, which is the
probability the qubits are in the initial state. If the probability
collapses and revives over time then we can conclude that the three
qubit system also exhibits `Collapse and Revival' of entanglement.
The results are shown in Fig. \ref{fig:reventangle3}. The measure
stated in Eq. \ref{eq:3qpuremeasure} can not be used here as the
qubits are mostly in a mixed state and Eq. \ref{eq:3qpuremeasure} is
for a pure state. For this reason we have chosen to look at the
probability of being in the initial state.

The qubits are started in a maximally entangled GHZ state,
($\tau_{ABC}=1$), that lies in the basin of attraction and are then
allowed to evolve. At a later time the entropy goes close to zero
and the qubits are close to the product form shown in Eq.
(\ref{eq:nqattract}). This is a completely unentangled state and as
the entropy is also zero in the large $\nbar$ approximation, there
is no entanglement in the entire system. After the `attractor' time
the entropy increases and is no longer close to zero. We follow the
green curve in Fig. \ref{fig:reventangle3} and note that
$P_{init}(t)$ increases to a value near 0.8 and therefore we propose
there is some three qubit mixed state entanglement present in the
qubit system. In the large $\nbar$ approximation this value does go to
unity, so the system is in the initial state, up to a phase. In
this approximation we will therefore necessarily get a full revival of
three-body GHZ entanglement.

As further confirmation that the entanglement is of 3-body
form, we have also calculated the entanglement present between just
two of the qubits after tracing out one of the qubits. It was found
that the tangle between any two qubits remained zero throughout the
evolution. Therefore we can conclude the entanglement present is
just GHZ entanglement and there is no W component present (as a W
component would lead to non-zero 2-qubit entanglement).

We have demonstrated the revival of entanglement for the numerically evaluated exact solution for two and three qubits, and shown it to be true for all
values of $N_q$ in the large $\nbar$ approximation. Of particular note is
that the type of entanglement we observe is of the GHZ form and is
thus a true $N_q$-qubit entanglement that cannot be reduced to a
simpler form. Remarkably, this implies that the quantum information
encoded in an $N_q$-qubit GHZ entangled state---for any value of
$N_q$---can be encoded into the state of a single field mode.


\section{Conclusion}

We have studied the `Collapse and Revival of Entanglement' in the
two, three and many qubit JCM. We have shown that the idea of an
`attractor' state discovered by Gea-Banacloche \cite{Ban90} for one
qubit does indeed exist in the multi qubit cases described by the
Tavis-Cummings model. Unlike the one qubit case, the qubits have to
be within a basin of attraction in order for the qubit system to
disentangle itself from the field at certain times. When the qubits
are started in a state in the basin of attraction then the system
behaves in a similar way to the one qubit case.

The basin of attraction contains mostly entangled states and there
are some interesting entanglement properties associated with the
whole system when the qubits are started in this basin of
attraction. The system evolves and there is then entanglement
between the field and the qubit system, but when the system goes to
the `attractor' state all entanglement has completely left the
system. The field and the qubits are both in pure states with the
qubit state being in the completely unentangled `attractor' state.
At a later time it has been found that the entanglement does return
to the system and then the process repeats. We represented explicit
examples of entanglement present between the qubits and also between
the qubit system and the field. We have called this phenomenon
`Collapse and Revival of Entanglement'. For initial qubit states
outside the basin of attraction, sudden death and rebirth
\cite{Eberly07} of entanglement can occur.
\appendix

\section{Spin Coherent States}\label{ap:spin}

There exist spin states analogous to the coherent state of the
harmonic oscillator. For the one dimensional harmonic oscillator the
coherent states are functions of a variable $\alpha$ which runs over
the entire complex plane:

\begin{eqnarray}
\ket{\alpha}&=&\frac{1}{\cal{N}}e^{\alpha a^\dagger}\ket{0}\\
&=&\frac{1}{\cal{N}}\sum_{n=0}^{\infty}\frac{(\alpha a^{\dagger})^n}{n!}\ket{0}\\
&=&e^{-\abs{\alpha}^2/2}\sum_{n=0}^{\infty}\frac{\alpha^n}{\sqrt{n!}}\ket{n}
\end{eqnarray}
where the normalisation factor $\cal{N}$ $ = e^{\abs{\alpha}^2/2}$.

These states from a complete set, in the sense that
\cite{GerryKnight}
\begin{equation}
\frac{1}{\pi}\int\ket{\alpha}\bra{\alpha}
d^2\alpha=\sum_{n=0}^{\infty}\ket{n}\bra{n}=\bf{1}
\end{equation}
where the right hand side is the unit matrix, or identity operator.

To find the analogue for spin half particles we consider a single
particle of spin $S=N_q/2$. We define the ground state $\ket{0}$ as
the state such that $\hat{S}^z=S\ket{0}$ where
$\hat{S}^z=\sum_{i=0}^{N_q}\sigma^z_i$ is the operator of the $z$
component of spin. Then the operator
$\hat{S}^+=\frac{\hat{S}^x+i\hat{S}^y}{2}$ creates spin deviations.
In a similar way to the field coherent state the spin coherent state
for many spin half particles is \cite{Radcliffe,Arecchi,Zhang}:
\begin{eqnarray}
\ket{z,N_q}&=&\frac{1}{\cal{N}}e^{z S^{+}}\ket{0}\\
&=&\frac{1}{\cal{N}}\sum_{n=0}^{N_q}\frac{(z S^{+})^n}{n!}\ket{0}\\
&=&\frac{1}{(1+\abs{z}^2)^{N_q/2}}\sum_{m=-N_q/2}^{N_q/2}\sqrt{\frac{N_q!}{(\frac{N_q}{2}-m)!(\frac{N_q}{2}+m)!}}z^{(\frac{N_q}{2}-m)}\ket{N_q,m}\label{eq:spincoh}
\end{eqnarray}
where the normalisation factor $\cal{N}$ $ =
\frac{1}{(1+\abs{z}^2)^{N_q/2}}$, $m=\sfrac{N_e-N_g}{2}$ is the spin
of the state and $z$ is a complex number that determines the center
of the coherent state. A spin coherent state is a product state and
therefore has no entanglement. For example if $z=1$, then the
average value of the spin coherent state is $\ket{N_q,m=0}$, if
$z=0$ then the average value of the spin coherent state is
$\ket{N_q,m=\sfrac{N_q}{2}}$ and if $z=\infty$ then the average
value of the spin coherent state is $\ket{N_q,m=-\sfrac{N_q}{2}}$.

The states $\ket{z,N_q}$ form a complete set and the completeness
relation is:
\begin{equation}\label{eq:completeness}
\frac{N_q+1}{\pi}\int\frac{\ket{z}\bra{z}}{(1+\abs{z}^2)^2}
d^2z=\sum_{m=-N_q/2}^{N_q/2}\ket{N_q,m}\bra{N_q,m}=\bf{1}
\end{equation}
In our work we present the `attractor' states for many qubits in Eq.
\ref{eq:Nqattractb} as spin coherent states. These states themselves
are indeed spin coherent states. If we set $z=ie^{i\theta}$ in Eq.
\ref{eq:spincoh} then we get $\ket{\psi^+_{N_q\,att}}$ and if we set
$z=-ie^{i\theta}$ then we get $\ket{\psi^-_{N_q\,att}}$.

When looking at the coherent state of a harmonic oscillator, there
exists a set of states called Schr\"{o}dinger cat states that are
the addition of two coherent states that are macroscopically
different, e.g. $\sfrac{1}{\sqrt{2}}(\ket{\alpha}+\ket{-\alpha})$.
There also exist spin coherent states that are analogous to these
Schr\"{o}dinger cat states. They are also defined as the addition of
two macroscopically different states, i.e.
$\sfrac{1}{\sqrt{2}}(\ket{z,N_q}+\ket{-z,N_q})$.

In our work we have plotted $Q$ function diagrams for the cavity
field state using the coherent state. There also exists an analogous
function for the qubits, defined in terms of spin coherent states in
a similar way, using the completeness relation Eq.
\ref{eq:completeness}. This spin $Q$ function is given by
\begin{equation}
Q_{q}(z,t)=\frac{\bra{z}\rho_q(t)\ket{z}}{\left(1+\abs{z}^2\right)^2}
\; .
\end{equation}

We have used this in our analysis to make interesting plots which
demonstrate pictorially that the states in the basin of attraction
for multi-qubit systems are indeed  Schr\"{o}dinger cat states of
the qubits.

\smallskip
\acknowledgments

The work of C.E.A.J. was supported by UK HP/EPSRC case studentship,
and D.A.R. was supported by a UK EPSRC fellowship. We thank W. J.
Munro for helpful discussions.

\smallskip

\bibliography{Dynamics_entanglement_in_TCM}

\end{document}